# Cosmology and general relativity in upper secondary school through new targeted teaching materials: a study on student learning and motivation

Authors[1]: A. Gasparini[a)], A. Müller[a),b)], F. Stern[a)] A, L. Weiss[a)]
Affiliations:
a) Institute of Teacher Education, University of Geneva, Switzerland
b) Department of Physics, University of Geneva, Switzerland

## Abstract

*Despite its intellectual and philosophical significance, cosmology and general relativity (GR) remain largely inaccessible to high-school physics teaching because of the advanced conceptual and mathematical prerequisites to master these topics. Integrating them into upper secondary physics teaching, outside specialized courses, poses a significant challenge that remains unresolved. While the strong rationale for their inclusion in the curriculum is widely acknowledged, significant obstacles, both practical and conceptual, hinder implementation. Moreover, empirical evidence on successful classroom and curriculum integration of modern physics remains scarce. This contribution reports on an implementation study of a GR and cosmology course developed for upper secondary school students as part of an educational project launched during the centenary of GR in 2015 and tested ever since for several years. The course aimed to expand students' knowledge to include current physics topics while highlighting their foundations in areas of classical physics such as Newtonian mechanics, electromagnetism, and waves. Targeted teaching and learning materials are focused on conceptual and qualitative understanding, while systematically combined with a mathematical treatment accessible at the upper secondary level, avoiding oversimplification. A key element is an active learning approach, incorporating activities and tasks such as engaging applications related to current research, reflective exercises, thought experiments, and hands-on tasks. The main research objective was to explore whether a conceptually deep and educationally effective GR and cosmology course could be successfully implemented for non-specialist upper secondary students. A pre-post study assessed both conceptual learning and affective outcomes, including interest, curiosity, self-concept, and perceived relevance of science. Results showed encouraging gains in both learning and motivation, with large to very large effect sizes for conceptual learning of core principles. Additionally, no or small effects of predictors such as gender were observed. We conclude that the integration of GR and cosmology into upper secondary physics teaching, in the form of courses and materials that are engaging, comprehensible, and impactful, is feasible. This can enable learners in non-specialist courses and with diverse backgrounds to engage with these intellectually and educationally rich and stimulating topics.*

## 1 Introduction

During the past century, numerous upheavals have occurred in the scientific world. Particularly, the contribution of the theory of general relativity (GR) has not only led to a revolution in our vision of the universe, but it has also brought forth numerous epistemological and philosophical questions that remain highly relevant today (see e.g. [1]; [2]; [3]; [4]). The universe at large serves as the ultimate laboratory for testing this emblematic theory, making cosmology the domain of choice for this purpose. In this regard, modern cosmology can be seen as the branch born from the relativistic view of nature. Indeed, over the past century, and especially in recent decades, significant observational advancements have been pivotal in transforming cosmology from an approximate branch of astrophysics into a highly precise science, situated at the intersection of the most advanced scientific knowledge and technologies [5].

[1] Author contributions (according to CRediT, https://authorservices.wiley.com/author-resources/Journal-Authors/open-access/credit.html): Conceptualization : A. G, A. M; Formal Analysis : A. G., M. D, F. S.; Investigation : A. G, A.M; Methodology : A. G., M. D, A. M., F. S.; Project Administration : A. G, A. M; Supervision : A. M.; Writing : A. G, A. M, with contribution by M. D, F. S.

Despite its intellectual and philosophical significance, the knowledge related to cosmology and GR remains inaccessible to the non-experts and the general public, mainly because of the level of mathematical and technical expertise required to master those topics, and the integration of those disciplines into schools remains a significant challenge that remains largely unresolved today. Nevertheless, the educational importance of astrophysics, and in particular cosmology, has been pointed out by recent recommendations from the National Academies of Sciences (USA) [6] highlighting "a clear benefit" by "capturing the public's attention with discoveries (...), promoting science literacy, and realizing advanced technologies that can then find real-world applications." More specifically, regarding secondary education, [6] emphasizes astronomy education "as a broad gateway to STEM careers" and highlights the potential of "embedding computational training". Indeed, many groundbreaking cosmology observations have been made in the last decades or are to be expected, such as by the LIGO interferometers, the James Webb Space Telescope, the Euclid satellite (in construction), and the LISA space interferometer and the SKAO project (planned) [7]. These observations and others are not only a reservoir of valuable data for fundamental research, but also an excellent opportunity to provide new and fascinating teaching content in schools.

Consequently, the integration of GR and cosmology into upper secondary physics teaching has been intensely discussed, with strong rationale on the one side, but also meeting considerable obstacles on the other. As Michelini states in her recent review [8], empirical work and evidence on successful classroom and curriculum integration of modern physics are scarce. Therefore, this contribution reports on an implementation study of cosmology and GR in an upper secondary physics course for non-specialists. The next section will present the discussion and research in this field more in detail.

## 2 Rationale and Research Background

### 2.1 The Value of Modern Physics in Upper Secondary Education

Teaching about relativity and cosmology is part of the larger perspective of integrating modern physics into secondary school physics (i.e. International Standard Classification of Education (ISCED) level 2.4.4 and 3.4.4; [9]). The case for the educational value of modern physics[2] has been made for decades ([14]; [15, sect. 1]), and Michelini [8] eloquently summarizes the current state of affairs, as "one of the main discussed content areas" in physics education, in particular regarding research-based teaching approaches on that matter.

One of the key arguments is its significance for the scientific worldview and its profound impact on fundamental philosophical concepts ([16, ch. 21]; [17, ch. 50.6]; [18, sect. 5]). As Levrini [19] states, modern physics "introduced radical modifications in human thought that every educated citizen should have at least heard of," making it an indispensable element of scientific literacy. Moreover, modern physics can help to overcome a "deformed vision of science" [20], providing motivation through insight into authentic research, historical breakthroughs, and access to current developments. It thus can add "a new quality to the curriculum, which urgently needs transformation as the physics taught in schools increasingly lags behind" [21]. In that sense, the implementation of modern physics in secondary schools is strongly advocated by both physics researchers and physics educators for areas such as quantum physics (e.g. [22]; [23], sect. I : "Teaching/Learning Quantum Physics"; [24]); relativity (both special and general, see e.g. [25]; [26]; [27]; [28]); particle physics (e.g. [29]; [30]; [31]; [32]); astrophysics and cosmology (e.g. [33]; [25]; [27]; [6]; [34] and references therein); and other fields (e.g. [23] sect. 5; [21];

[2] "Modern physics" is commonly understood ([10]; [11]) to start with the advent of quantum mechanics and relativity at the beginning of the 20th century, and it is put in contrast to classical physics, fundamentally challenging several of its core concepts and assumptions. Some researchers also include statistical physics based on the then recent evidence for the existence of atoms ([12], p. 2), or extend the begin of modern physics the discovery of X-rays in 1895 [13].

[35]). Work e.g. by Michelini [8] or the initiative IMPRESS (International Modern Physics and Research in Education Seminar Series, [36]; [37]) provide an overview of current work.
However, as a pioneer of physics education reform already deplored [38], the "sad fact" is that the very abstract level of modern physics topics and "the remoteness of their models from common experience make them exceedingly difficult for laymen to grasp". Of course, the advanced mathematics needed is a particularly difficult element of this deplored abstractness. Thus, for the implementation at school to work, one needs "productive forms of complexity" [39]. Yet, as Michelini underlines in her recent review [8], empirical studies and evidence regarding the successful integration of modern physics into upper secondary physics teaching are limited, and little is known about how to achieve "productive forms of complexity" within the practical constraints of a typical physics classroom setting. In particular, there is an urgent need to avoid limited approaches that provide no more than notions in an informal or narrative way. As mentioned above, this work presents a study on the implementation of cosmology and GR in an upper secondary physics course designed for non-specialists. The following sections provide the research background and specify the research objectives of the study.

### 2.2 Effects on learning and motivation

From the learning perspective, results of large-scale studies have shown that students after having taken courses on modern physics significantly improve their learning, including topics of classical physics such as Newtonian mechanics or electromagnetism, because of the reactivation of the necessary background in the application to the new context ("cumulative learning" [40]). Stephen Weinberg has provided a very nice example of this necessity of classical underpinnings for the understanding of modern physics in his popular book on elementary particle physics [29]. Note, however, that other research points out how excessive simplification, aimed at making topics of modern physics more accessible, can also have negative effects on students' understanding [41].
Additionally, studies in science education also demonstrate the interest of astronomy and cosmology in enhancing learners' understanding of the nature of science and its specifics (NOS). By NOS, we mean, in line with [42], everything related to the role of hypotheses, observation, experimentation, refutation, etc., which often synthesize into the approach advocated by the "scientific method".
Regarding motivation, there is strong evidence indicating that, among all scientific disciplines, astrophysics and cosmology are perceived as among the most interesting by young people, with this perception being equally shared between girls and boys [43]. The gender issue has an important role because, despite political efforts aimed at change, the proportion of women pursuing careers in science, technology, engineering, and mathematics (STEM) fields, particularly in the "hard" sciences and technical domains, remains consistently low across all countries internationally [44]. This is true not only at the academic level, but also for math and physics options at upper secondary II level (e.g. the percentage of women in such courses in Switzerland is below 25% [45]).
In addition, addressing topics directly connected to the academic research in physics allows students to:

- Expand their knowledge to include current areas and questions in physics that go beyond the traditional upper secondary curriculum. Indeed, although secondary school programs traditionally taught in other scientific branches (such as biology or chemistry) have evolved in response to recent advances and incorporated developments from the past century or decades, the latest progress in physics cannot be taught without covering the basics – many of which remain the same as they were a century ago. Moreover, these fundamentals (like Newtonian physics, electromagnetism, or waves) form the basis of almost all scientific and technological knowledge that serves society today.

- Become aware of the role and robustness of the physics they learn in basic courses. Although these courses may not cover the most advanced theories, they provide models with impressive accuracy for a wide range of natural phenomena, from the molecular scale to astrophysical scales.

For these reasons, we believe that this course can positively influence students' perception of the importance of the physics course, both in terms of personal orientation and within a broader societal context.

Finally, we mention how topics related to astrophysics, especially cosmology and fundamental physics, have an inherent capacity to evoke a range of sensations and emotions that have resonated with humanity since ancient times. These emotions encompass wonder, elevation, sublimity, and a sense of apprehension. This emotional state, referred to as "awe", has been recently acknowledged for its potential to positively influence affective outcomes and, by extension, learning outcomes ([46]; [47]). Besides, awe is typically experienced by astronauts when they view the Earth from space and is typically associated with what is known as the "overview effect" [48]. This includes a heightened awareness of the interconnectedness and vulnerability of our planet, leading to a relativization of human-centric perspectives in light of the immense scale of nature and fostering a humbler perception of the scientific approach. More generally, cosmology and fundamental physics are fields that do not exclusively emphasize the utilitarian practice of science, but rather reassess its philosophical and epistemological aspects. This also amounts to highlighting the aesthetic and poetic essence of science, more than its prevalent prosaic character in our society ([49]; [50]). This appreciation of scientific research is highly relevant in the current debate regarding the role of science in a transitioning world.

### 2.3 Research questions

Given the international recognition of the importance of teaching cosmology and GR, many initiatives have recently emerged with the aim of introducing these disciplines into secondary schools. In several countries, specific contents have already been incorporated into the official curriculum, notably in Austria, Estonia, Norway, and Sweden [51]. Moreover, the richness and potential benefits of achieving even a conceptual and qualitative understanding of these subjects have motivated educational projects in this direction, for instance, initiatives in Australia, Sweden [52], Germany [53] and Italy [54].

However, as should be clear from the research background presented above, the primary limitation in introducing cosmology and GR into schools remains that a rigorous treatment of these subjects demands mathematical skills and knowledge that easily exceed the typical curriculum at secondary school and even at early university levels.

Regarding in particular the situation at the upper secondary level, the necessary educational effort is notable to produce fairly in-depth educational material without it becoming technically inaccessible, while still maintaining the educational and cultural benefits of a deep enough and global vision of these subjects. Very few such teaching resources exists in the present day, despite the strong rationale and the tangible demand for it as expounded above. This study thus aims to contribute to the development of such teaching resources and provide evidence addressing the following research questions:

RQ1: To what extent is it possible to develop and deliver a cosmology and GR course at the upper secondary school level that is technically deep enough to fully exploit the physical and philosophical potential of the target group?

RQ2: What is the impact of such a course on students learning of cosmology and general relativity?

RQ3: What is the impact of such a course on affective variables such as interest, curiosity, self-concept, and relevance of science?

# 3 Description of the intervention

The intervention of this study was based on a course of cosmology and GR developed starting from an educational project launched by the SwissMAP research hub (https://nccr-swissmap.ch/) in 2015, on the occasion of the centenary of GR. The funding for this project continued until 2017, thereby enabling the creation of teaching materials forming the basis of a course in these subjects for upper secondary students, which is still being utilized and further developed today.

## 3.1 Framework of the project

The SwissMAP project was launched with the following objectives.

- Provide students of the upper secondary with a relatively complete and in-depth understanding of issues related to cosmology and GR, as well as certain aspects of modern physics. To achieve this, it was necessary to create a course, as existing educational resources on these topics were either basic and geared towards the general public or primary and early secondary school students, focusing on qualitative and philosophical aspects (characterized by the expression "zero equations"), or at the "expert" level, for students in higher university cycles.

- Strengthen and broaden students' knowledge in physics and mathematics through fascinating and motivating subjects for them. The goal is not to replace and strengthen traditionally taught subjects in physics and mathematics courses, but to complement them by delving into their applications in cosmology.

- Improve the connections between secondary school and the world of research: students studying, as is traditionally done, 19th century physics often have biased ideas about the main issues and approaches of modern physics. A better awareness of contemporary research in physics (and thus the study options available to students after their secondary school diploma) can have significant implications for young people's career orientation choices. The aim was therefore to contribute to a general positive impact on the interest in the hard sciences.

The course is based on educational material consisting of a book [55] with eight main chapters, ranging from the introduction to astrophysics to gravitational waves, including gravitational lensing, black holes, and cosmological equations, among others. Seven appendices integrate and deepen complementary concepts that students may need for a smooth understanding of the main course, such as the tidal effect, blackbody radiation, or the tunneling effect. Each chapter includes a series of related exercises [56] freely available online with answer keys reserved exclusively for teachers[3]. It was thereby implemented in the classes starting from 2016 and, since the start of the 2018 school year, the course, originally written in French, has been successfully delivered in at least a tenth of classes in Switzerland (including Italian- and German-speaking students), and other students in France, Belgium, and Canada (in French), with continuous feedback and input from students and teachers. Additionally, many experts in cosmology, GR, and physics education at University of Geneva have provided their remarks and comments, which were considered during the publication phases in 2018 and its French and Italian (and English) re-editions in 2024. Improvements and extensions have been made since 2018 and the exercises created each year contribute to the existing repertoire. Indeed, questions that arise based on further deepening a subject or evolving scientific developments provide opportunities to add new explanations. This has been the case, for example, with observations of high-redshift lenses and galaxies by the James Webb Space Telescope, or with images of supermassive black holes by the Event Horizon Telescope collaboration.

[3] https://nccr-swissmap.ch/school-teachers-children/general-relativity

### 3.2 Swiss school structure and optional courses

In the Swiss secondary education system students interested in the sciences can choose a specific option (SO) that may be physics and applications of mathematics (PAM), corresponding to the "specialization" in their maturity. All other students follow only two years of physics at a level referred to as "fundamental discipline" (FD), where they study standard physics based essentially on concepts discovered by physicists until the 19th century. To expand the course offerings in the fundamental discipline of physics, students who had **not** chosen the specific option PAM are offered the opportunity to study physics in a complementary option (CO) course spanning four semesters, based on modern physics, to account for the advancements of the 20th century. The course of cosmology and GR was integrated into the curriculum for a semester as a component of this physics CO course.

We emphasize that this CO is not a specialized physics course, but it is rather a choice offered to the students who, upon entering the gymnasium, have chosen a specialization (SO) in disciplines **other** than physics (languages, arts, biology-chemistry, economics-law, etc.) and have the opportunity to complete their curriculum by taking a path aiming to broaden their education to other fields than their specialization. The CO course has a weekly allocation significantly less than the specialized courses, and its main objective is to improve general scientific knowledge, more than to provide advanced knowledge. On one hand, those students possess the basic concepts of physics and mathematics from previous FD courses, which are essential for approaching a subject like cosmology. On the other hand, they are open to learning new content – different from that of their specialization – which can influence their future university choices.

### 3.3 Course content

Here, we shortly present the course content covered with the classes participating in this study, mainly concerning the first five chapters. A more detailed presentation of the possible sequences can be found in reference [57].

#### 3.3.1 From astrophysical introduction to the accelerated expansion

Given that beginner students have almost no knowledge of astrophysics, the course begins with an introduction to the composition and structures of the universe: from the smallest to the largest distance scales. This panoramic allows for an initial awareness of the orders of magnitude involved and of the key observational elements developed in the last century and revisited later in the course, such as the chemical composition of baryonic matter, the homogeneity of structures at large scales, or observations of the galaxies' rotation curves and their velocities in clusters. The latter observations are the basis for the existence of the hypothetical dark matter, composed of particles interacting gravitationally but not electromagnetically. This initial chapter provides an opportunity to revisit and develop various basic concepts, such as Newton's laws in the context of circular motion, free fall, and the nature of gravity – comparing it with electromagnetic interaction. It also involves estimating orders of magnitude and unit conversions.

The second chapter covers mostly the elements related to the kinematics of the universe. After analyzing the difference between the Doppler effect and cosmological redshift, and its connection with the conceptual considerations of the new vision of Einsteinian gravity, the Hubble-Lemaître law and the concept of the rate of expansion of space are introduced. Through unit conversions, students formulate the Hubble-Lemaître law at different distance scales and notice that, although the expansion velocity can indeed be largely neglected at human-scale distances, it cannot be ignored at cosmological scales. Here is, among other, also an opportunity to focus on one hand on the question of the limit of the speed of light in a vacuum, i.e. understanding when this speed is a limit (motion of bodies *in* the space) and when it is not (motion of the space itself), and, on the other hand, on the dimensionality of the Hubble constant and the concept of characteristic time, by comparing it to the durations of different phenomena. The observation of the universe's expansion in the early 20th century was crucial for the new vision of the cosmos and was the starting point for other predictions and observations that further supported the Big Bang model and the underlying new vision of gravity. Examples include the cosmic microwave background (CMB) and the chemical abundances of light elements. This understanding persisted until the end of the last century when observations of Type Ia supernovae disrupted this coherent view. These observations revealed the recent acceleration of expansion, leading to the introduction of a hypothetical "dark energy" as its explanation. This is an opportunity to introduce the idea of the cosmological constant in modeling the dynamics of the universe, mentioning its edifying history: initially introduced by Einstein before the observation of expansion (to make his equations compatible with a static universe), it was later regretted and then reintroduced at the beginning of the 21st century to model acceleration. The chapter on expansion concludes with a summary of the proportions of the main mass-energy components of the observable universe. The topics covered in the first two chapters provide an opportunity to enhance the course with activities and exercises using images and data from satellite telescopes like Hubble or James Webb (JWST). In these activities, various cosmological redshifts and distances can be calculated based on the absorption lines of different chemical elements (Fig. 1).

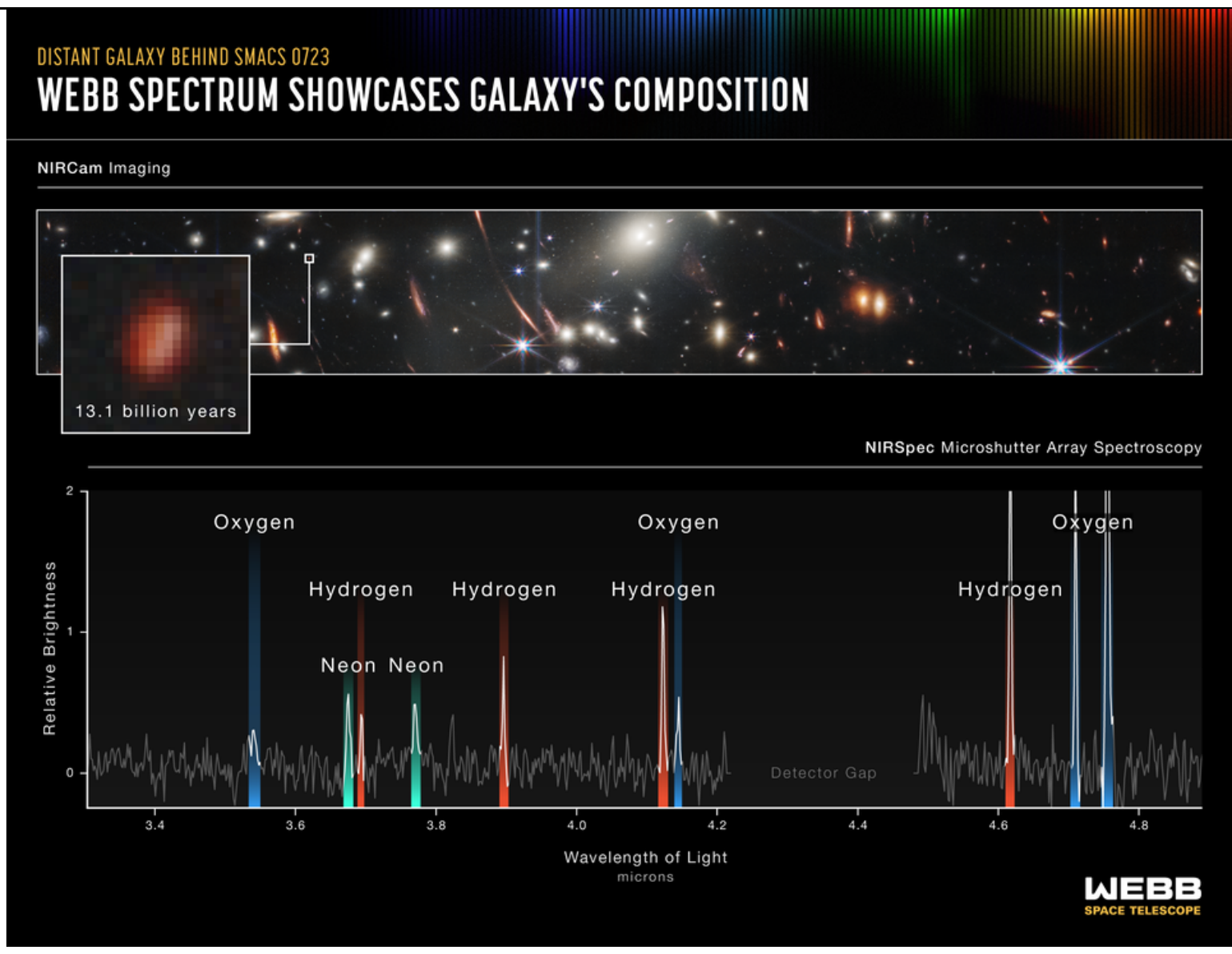

Fig. 1. Students calculate the cosmological redshift of the source (red in the enlargement) based on the redshift of the emission lines of chemical elements, and deduce whether it is beyond the Hubble radius and explain why it can be observed even if the expansion velocity exceeds the speed of light in a vacuum.

### 3.3.2 Basics of general relativity

The first two chapters of the course motivate the introduction of GR concepts by revealing the need for a new understanding of gravity in describing the universe. In Chapter 3 of the course, the characteristics of gravitational interaction are explored in order to introduce the equivalence principle, the keystone starting point for developing Einstein's new vision of inertial motion and of the nature of space-time. The concept of free fall and the definition of equilibrium are entirely reconsidered from a broader perspective than in basic physics course ("fundamental discipline"), allowing for a genuine rediscovery of these concepts by students. The concepts of Gaussian curvature, total curvature, or geodesics are then explained, always based on upper secondary mathematics, using examples in two or three dimensions while ensuring that their original conceptual scope remains intact. For instance, this includes analyzing the two-dimensional equivalent of the emblematic case of space curvature and the behavior of geodesics around a mass/energy concentration with spherical symmetry (Fig. 2). Students can independently verify, using a ribbon or tape on a three-dimensional plastic, the behavior of parallel geodesics on parts of the surface with positive or negative curvature.

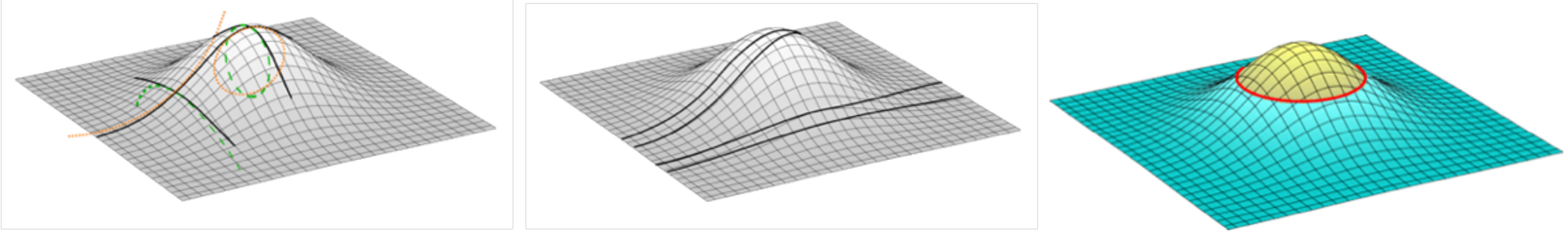

Fig. 2. The curvature at a point on a surface can be determined in two ways: by comparing the signs of the curvature radii of osculating circles in the principal directions (left) or by the behavior of parallel geodesics (center). For a mass concentration with spherical symmetry, the central area has positive curvature, and the peripheral area has negative curvature (right).

### 3.3.3 Gravitational lensing and black holes

The next two chapters, namely the fourth on gravitational lensing and the fifth on black holes, focus on two remarkable applications of GR. They constitute the second part and the highlight of the course. Here, the application of contents from the previous chapters as well as physics and mathematics of the traditional courses is leveraged to achieve a conceptual and quantitative understanding of these topics. Although the mathematics of GR, underlying these models, are far too advanced to develop a rigorous approach to these subjects with upper secondary students, we know that GR is linked to Newtonian mechanics in that it is a generalization thereof, often adding corrections to the Newtonian treatment, which we will introduce later, justifying them qualitatively. For the sake of readability of this article, which aims primarily to present the research results, we will not continue the detailed description of the concepts, applications, and methods deployed in these two chapters, referring the interested reader to the explanations in Supplemental Materials [58] and reference [57].

# 4 Methods

## 4.1 Sample and Setting

A total of 70 students took the cosmology course as a semester module in a complementary physics option (CO) during the academic years 2020-2021 to 2023-2024. This included four classes of students aged between 18 and 19 years. The courses were taught by the same teacher. In Tab. 1, details of the enrollments are provided. It is noteworthy that this course is the only optional physics course with a gender ratio (number of females: number of male participants) ≥ 1, while the average for this parameter across Switzerland is around 0.3, reflecting the same average in the early years of hard science faculties (physics, mathematics, computer science, etc.).

To encourage a reliable completion of the questionnaires, an explanation about the purpose of improving teaching quality on the basis of the gathered data was provided to the students at the beginning of the course. A conceptual (sect. 4.3.1) and an affective (sect. 4.3.2) questionnaire were administered as pre- and post-test, and completed by 68 and all 70 students, respectively. The conceptual post-test contributed to the participants' semester grades.

Tab. 1 Sample of the study: 4 cohorts of students who took the semester-long cosmology complementary option (CO) course between 2020 and 2024. The decrease in the last cohort is due to the fact that the course enrolment was capped by the school. $N_F$ et $N_M$ correspond to the respective number of girls and boys. For comparison, the gender ratio $N_F$ / $N_M$ is also provided for the specialized physics courses and its average through the swiss canton.

| School year | N | $N_F$ | $N_F$ / $N_M$ complementary option (CO) | $N_F$ / $N_M$ specialized physics courses (SO) | $N_F$ / $N_M$ in the high school |
|---|---|---|---|---|---|
| 2020-21 | 16 | 8 | 1.0 | 0.28 | 1.3 |
| 2021-22 | 22 | 12 | 1.2 | 0.30 | 1.3 |
| 2022-23 | 20 | 12 | 1.5 | 0.33 | 1.2 |
| 2023-24 | 12 | 6 | 1.0 | 0.27 | 1.2 |
| Total | 70 | 38 | 1.2 | 0.30 | 1.3 |

Each cohort completed the physics CO course over a total of 2 years (4 semesters). The cosmology course was one of the four modules included in this course, specifically scheduled for the first semester of the second year. It was preceded by a module on radioactivity (first semester of the first year) and a module covering additional basic physics concepts (ballistics, solid body mechanics, and fluids) during the second semester of the first year. Following the cosmology module, students took a module on electricity and magnetism and their applications (second semester of the second year).

### 4.2 Design and Timeline

Each year, the cosmology course lasted for a semester – from August to January – comprising 16 lessons of 2 periods of 45 min each and a final written exam. The first and last lessons were dedicated to the administration of pre and post-tests, respectively, with the aim of assessing the impact of the course on conceptual knowledge in cosmology (RQ2) and affective variables (RQ3).

Additionally, semi-structured interviews were conducted following the post-tests, between 3 and 6 months after the course's completion, involving 15 students in total. Each interview lasted around 1 h. They began with questions about the student's choices of SO and CO courses ("Why did you choose these courses?"), followed by inquiries about their future plans ("What will you study in the future?"), and continued with questions about their CO lessons. These interviews have allowed, among other things, to complement and guide us in the interpretation of the quantitative results of the conceptual and affective tests, as well as to receive qualitative feedback from students on this newly implemented course.

### 4.3 Variables and Instruments

To assess the dependent variables, the following questionnaires were created, and the results were analyzed using standard item analysis [59].

#### 4.3.1 Conceptual learning

A questionnaire evaluating the impact of the course on conceptual knowledge in cosmology (RQ2) was administered in pre and post-tests to the participants. This conceptual questionnaire consists of 14 multiple-choice questions (MCQ) on the content of the course and had to be completed by participants in 20 minutes. Each question has only one correct answer among 5 choices, and one point was assigned for each correct response. In case of an incorrect answer or no response, no points were assigned. Since no conceptual questionnaire on cosmology at the upper secondary level existed, the conceptual questionnaire was created in 2018 by the authors (in French) and was subsequently improved in the two following years until its final form at the beginning of the study. It did not undergo further changes during the entire study period. The questionnaire, which is available as Annex document n. 1, was designed as an instrument to assess general knowledge across various topics in cosmology, without the intention of creating multiple sub-dimensions, each containing several items evaluating the same concept. In this context, each item can be considered independent while contributing to a test that evaluates general knowledge in cosmology. Thus, the questions assess various concepts, which can be categorized into three main domains covered during the semester:

- Astrophysical introduction on scales, composition and expansion of the universe (section 3.3.1) corresponding to items 1, 2, 3, 5, 6, 7, 8, 9, 14;
- Gravitational lensing, corresponding to items 4 and 10 (section 3.3.3);
- Black holes, corresponding to items 11, 12 and 13 (see. Supplemental Materials [58]).

#### 4.3.2 Affective variables

An affective questionnaire was administered in pre and post-tests to assess the impact of the course on affective dependent variables "self-concept", "interest", "curiosity state" regarding the course of the last semester, and the perception of "relevance of science" (RQ3). This questionnaire (in French, Annex document n. 2) contains 28 statements, where each student expressed their agreement on a scale ranging from 1 (strongly disagree) to 6 (strongly agree). For negated items the scale was inverted for the statistical analyses. The instruments were taken [60], in turn based on a selection of existing and well-validated scales (except for *relevance for science*, RS):

- *Interest* (IN), adapted from [61], [62] and [63];
- *Self-concept* (SC), adapted from [61], [62], [63] and [64];
- *Curiosity state* (CS) regarding the content of the physics course, adapted from [61], [62] and [63];
- Perceived *relevance for science* (RS) of the physics course was an additional variable that we deemed relevant.

Sample items and instrument characteristics are provided in Tab. 5.

#### 4.3.3 Predictor variables

Three potentially predictors related to learner characteristics were considered:

- *Gender*;
- *Prior physics knowledge, measured as prior grade as proxy* (phys_grade_pre);
- *Curiosity as a personality trait* (CT) was measured in the affective pre-test as a predictor, in addition to the independent variable *curiosity state* related to the content of the physics course (CS), using well-validated instruments is available from prior research in the field [60] adapted [65] and [66];

Additionally, it was tested whether a class (see Tab. 1) had a significant effect on learning or affective variables.

#### 4.3.4 Interviews

A series of 11 interviews were conducted with students who had taken the CO course of cosmology, as well as 4 interviews with students who had taken part in the cosmology course within the framework of their SO physics. These interviews took place each year from 2020 to 2024, between a few weeks and a few months following the end of the course.

The choice was made to conduct semi-structured interviews [67], which allowed the respondents to express themselves quite freely, adding details to their answers while staying within the intended framework. Thus, after gathering some personal information about the interviewed student (gender, chosen school options, academic results in mathematics and physics), they were questioned about the reasons for their choice of school options, and then about their future plans, since these students would obtain a high school diploma giving them access to tertiary education. A few questions aimed to make the student express (and realize) if and to what extent the cosmology course has allowed them to deepen their previous knowledge in mathematics and classically taught physics. Then, the door opened to their interest in science in general, modern physics in particular, and cosmology. The students were also asked about the topics covered in the cosmology course that they liked or disliked, as well as the quality of the course materials. Some took the opportunity to talk about the pedagogical qualities of the teachers, the articulation of the different modules of the CO physics course, and the particular management of the cosmology course. When the interview deviated from the main topics, the student was gently steered back to the main theme.

The interviews were recorded with the student's consent, and the interviewer simultaneously took notes. Once formatted, these notes are sent to the interviewee for validation. Out of the 15 interviews, only 2 people proposed modifications, one regarding a factual element and the other concerning the phrasing of a sentence, while the others approved the notes as accurately reflecting what they had said and intended to express.

### 4.4 Data Analysis

#### 4.4.1 Conceptual Learning

*Psychometric indicators*

For the newly developed conceptual learning questionnaire an item analysis was carried out according to standard procedures and requirements (see [59]), and calculating the following set of instrument characteristics for both the pre-test and post-test: item difficulty (*P*), standard deviation (SD), *discrimination index* (*D*), item-test correlation ($r_{it}$), and Cronbach's alpha if item deleted ($\alpha^*$, as well as $\alpha^*$-$\alpha$) in order to assess the impact of each item on the internal consistency of the scale, indicating whether removing an item would increase or decrease the overall internal consistency. On the instrument level the mean score across all items, range, and standard deviation are reported, as well as Cronbach α as measure of internal consistency.

Effect sizes are calculated as Cohen *d*, using the usual effect-size levels (small ($0.2 < d < 0.5$), medium ($0.5 \leq d < 0.8$), or large ($0.8 \leq d$; [68]). Another reference value of $d = 0.4$ is used by Hattie [69] as a "hinge point" between influences of smaller and larger size[4].

*Statistical analyses*

For each of the 14 items, we calculated the percentage change in correct answers from pre-test to post-test, as well as the percentage change in answers reflecting misconceptions from pre-test to post-test.

A repeated-measures *t*-test with Benjamini-Hochberg adjustment for multiple testing [70] was conducted to analyze the pre-post-test changes in the overall score. Potential predictor effects were tested through an ANCOVA (see 4.4.3).

#### 4.4.2 Affective variables

For the four affective dependent variables (*interest* (IN), *self-concept* (SC), *curiosity state* (CS), and *relevance for science* (RS); see 4.3.2) we calculated the mean score, range of values, item-test correlation, and Cronbach's alpha [71] for both pre-test and post-test, according to standard procedures for multiple choice questionnaires (see [59]); as for learning, pre-post changes were tested through a repeated measures *t*-test with Benjamini-Hochberg adjustment for multiple testing [70]. Cohen's *d* was computed for each item and the full instrument to assess effect sizes. The same set of instrument characteristics was calculated for the predictor variable curiosity trait (4.3.3). Potential predictor effects were tested through an analysis of covariance (ANCOVA, see 4.4.3).

#### 4.4.3 Predictor effects

An ANCOVA analysis was conducted to test for potential predictor effects: The analysis was performed for the variation of (i) learning outcomes, specifically conceptual understanding, and (ii) affective outcomes (i.e., self-concept, interest, perceived relevance of science, as well as curiosity state), with class, gender, curiosity as trait of personality, and prior physics knowledge as predictors.

Statistical significance was set at $p < 0.05$. All analyses were conducted with the R software [72].

## 5 Results

### 5.1 Conceptual Learning (RQ1 and RQ2)

#### 5.1.1 Item analysis

Tab. 2 gives an overview of the instrument analysis, including the average values and the standard deviations calculated on the values of the 14 items, as well as the ranges of values and the alpha.

[4] We agree with Hattie [69] that these thresholds are an element of discussion to be used with circumspection, not a value to be applied blindly. Lower effect sizes might well be worth considering, depending on available alternatives, effort, and so on, and vice versa for higher effect sizes.

Tab. 2. Instrument analysis of the learning pre- and post-test (14 items, N = 68). * Standard deviation is calculated on the average value of each of the 14 items.

| **Scale characteristic (SD*) [range of values]** | **Mean value (SD*) [range of values]** | |
|---|---|---|
| | PRE | POST |
| Item difficulty *P* | 0.35(0.18) [0.06 ; 0.60] | 0.76(0.15) [0.43 ; 0.93] |
| Cohen's *d* | 2.78 | |
| Item discrimination *D* | -0.01(0.29) [-0.51 ; 0.49] | 0.44(0.15) [0.18 ; 0.70] |
| Item-test correlation $r_{it}$ | 0,44(0.11) [0.14 ; 0.61] | 0.38(0.11) [0.23 ; 0.53] |
| Internal consistency *α* | 0.37 | 0.56 |

As we highlighted in section 4.3.1 on methodology, the conceptual questionnaire was designed as an instrument to assess general knowledge across various topics in cosmology, without the intention of creating multiple sub-dimensions, each containing several items evaluating the same concept. Thus, each item can be considered independent while still being part of an overall test evaluating general knowledge in cosmology. Therefore, we have no specific expectations regarding of value for the internal consistency of the test. Furthermore, the alpha value obtained in the post-test results is close to 0.6 and significantly higher than the alpha of the pre-test. Since this article aims to address the research questions posed in section 2, we will proceed with the test results in the next section. Focusing on the details of each item in the questionnaire, Tab. 3 presents the characteristic psychometric indices from the item analysis. The English translations of keywords (originally in French) are also provided.

The fact that the psychometric parameters of the pre-test do not meet the recommended ranges is unsurprising, considering that the majority of responses are incorrect (indicating "false" for any given response). Indeed, all parameters become satisfactory to very good levels in the post-test, and we observe a significant improvement across all parameters between the pre-test and the post-test.

Moreover, in the pre-test, for 10 items out of 15 the wrong responses were not evenly distributed, and the greater proportion of responses corresponded to one incorrect answer, revealing potential misconceptions. For each of those 10 items, Tab. 4 shows the percentage of right answer and of answers potentially corresponding to a misconception in the pre-test and in the post-test.

### 5.1.2 Learning gains

The repeated measures with adjustment for multiple measurements (sect. 4.4) yielded for the individual items statistically highly or very highly significant effects with effect sizes according to Tab. 3. For the overall score, a statistically highly significant effect with a very high effect size was found ($t = 20.3$, $p < 0.001$; $d = 2.78$).

ANCOVA analyses were conducted to examine the effects of class, gender, and prior physics grades on learning outcomes (see 4.3.3). Prior physics grades significantly influenced the outcome ($F(1, 62) = 10.90$, $p = .002$), while class ($F(3, 62) = 2.68$, $p = .054$) and gender ($F(1, 62) = 3.15$, $p = .081$) were not statistically significant.

## 5.2 Affective variables (RQ3)

### 5.2.1 Item analysis

The instruments for four affective dependent variables *interest* (IN), *self-concept* (SC), *curiosity state* (CS), and *relevance for science* (RS) concerning the CO module followed the last semester, plus the covariate *curiosity as a trait* of the personality (CT), were analyzed according to standard procedures and requirements [73], [59], [74]. For each variable, the analysis of the instrument, including the total mean values and the range of values of the characteristic quantities is shown in Tab. 5. We observe that all test properties are within the recommended ranges, attaining satisfactory to good values. The values of the pre-post Cohen's *d* are given as well for each affective output.

Tab. 3. Item analysis of the learning pre- and post-test (14 items, N = 68). Effect size: * = medium; ** = large.

| N | Item keywords (translated from French) | *P*(SD) | | Cohen *d* PRE/POST | D | | $r_{it}$ | | $\alpha^* - \alpha$ |
|---|---|---|---|---|---|---|---|---|---|
| | | PRE | POST | | PRE | POST | PRE | POST | POST |
| 1 | The Big Bang is a state of the primordial universe | 0.10(0.31) | 0.59(0.50) | 1.59** | 0.24 | 0.51 | 0.26 | 0.29 | **-0.017** |
| 2 | We can observe other galaxies moving away because… | 0.37(0.49) | 0.76(0.43) | 0.82** | 0.49 | 0.35 | 0.41 | 0.34 | +0.006 |
| 3 | Does the universe have a center? | 0.21(0.41) | 0.69(0.47) | 1.19** | 0.28 | 0.50 | 0.39 | 0.37 | +0.008 |
| 4 | Two parallel rays of light pass near a star. Among the following diagrams… | 0.60(0.49) | 0.85(0.36) | 0.51* | 0.11 | 0.50 | 0.16 | 0.55 | +0.057 |
| 5 | We call "dark matter" matter ... | 0.31(0.47) | **0.93**(0.26) | 1.33** | 0.29 | **0.18** | 0.50 | 0.23 | +0.003 |
| 6 | A ray of light leaving a galaxy at 700 ly takes +/- than700 years … | 0.60(0.49) | 0.87(0.34) | 0.54* | **0.15** | **0.28** | 0.38 | 0.22 | **-0.006** |
| 7 | The "cosmic microwave background" is the radiation … | 0.25(0.44) | **0.91**(0.29) | 1.52** | -0,07 | 0.32 | 0.34 | 0.39 | +0.025 |
| 8 | What is some evidence that the universe was very hot and dense in the past? | 0.44(0.50) | 0.76(0.43) | 0.65* | 0.02 | 0.45 | 0.35 | 0.40 | +0.021 |
| 9 | By "dark energy" we mean … | 0.31(0.47) | 0.87(0.34) | 1.20** | 0.04 | 0.50 | 0.43 | 0.52 | +0.050 |
| 10 | An observer aligned with two galaxies can see an image … | 0.43(0.50) | 0.65(0.48) | 0.44* | -0.16 | 0.35 | 0.28 | 0.28 | **-0.016** |
| 11 | A BH 100 times more massive than the other is 100/10000 times more/less dense. | 0.06(0.24) | 0.43(0.50) | 1.55** | -0.51 | 0.70 | 0.08 | 0.53 | +0.050 |
| 12 | The signal emitted by an object in a BH can reach an external observer … | 0.59(0.50) | 0.84(0.37) | 0.50* | -0.32 | 0.38 | 0.27 | 0.41 | +0.027 |
| 13 | Can an isolated black hole in the universe lose mass? | 0.16(0.37) | 0.60(0.49) | 1.19** | -0.31 | 0.74 | 0.38 | 0.53 | +0.052 |
| 14 | What is the best estimate of the age of the universe? | 0.43(0.50) | 0.84(0.37) | 0.83** | -0.36 | 0.35 | 0.31 | 0.34 | +0.012 |
| Total | | 0.35(0.15) | 0.76(0.16) | 2.78** | | | | | |

| Tab. 4. For 10 items out of 14 of the conceptual learning test, the pre and the post-percent of right answers is indicated in the white cells, whereas the same percent of answers associated with a potential misconception appears in the gray cells. | | | | | |
|---|---|---|---|---|---|
| **N** | **Item keywords (translated from French)** | **Right answer** | **Percent of right answers PRE → POST** | **Related misconception** | **Percent of answers related to the misconception PRE → POST** |
| 1 | The Big Bang is | … a state of the primordial universe that current physical theories cannot describe. | 10% → 59% | … an explosion… | 79% → 25% |
| 2 | We can observe other galaxies … | … moving away more quickly as their distance is greater. | 37% → 76% | … moving away with the same speed/almost at rest. | 46% → 24% |
| 3 | Does the universe have a center? | No. | 21% → 69% | We don't know. | 60% → 21% |
| 5 | We call "dark matter" ... | matter which does not interact electromagnetically | 31% → 93% | … particles of antimatter. | 37% → 0% |
| 7 | The "cosmic microwave background" is the radiation … | …emitted in the primordial universe when neutral atoms were formed. | 25% → 91% | … emitted during the formation of the first galaxies. | 37% → 6% |
| 8 | What is some evidence that the universe was very hot and dense in the past? | The chemical composition of the universe is currently about 75% hydrogen and 25% helium. | 44% → 76% | The measure of the density of the universe. | 31% → 21% |
| 9 | By "dark energy" we mean … | … the energy at the origin of the change in volume of empty space. | 31% → 87% | … the energy of the dark matter. | 41% → 3% |
| 11 | A BH 100 times more massive than the other is | 10000 times less dense. | 6% → 43% | … the same / 100 / 10000 times denser. | 71% → 25% |
| 13 | Can an isolated black hole in the universe lose mass? | Yes, if its temperature is high enough. | 16% → 60% | We don't know. | 46% → 10% |
| 14 | What is the best estimate of the age of the universe? | About 15 billion years. | 44% → 84% | About 150 billion years. | 34% → 9% |

### 5.2.2 Affective variables gain

We observe an increase in the averages for *curiosity state* ($t(69) = 3.34$, $p = 0.001$, $d = 0.46$) and self-concept ($t(69) = 1.72$, $p = 0.090$, $d = 0.26$), as well as for interest (however not statistically significant, $t(69) = 1.41$, $p > 0.05$, $d = 0.17$); no change for perceived relevance of science was found ($t(69) = 0.05$, $p > 0.05$, $d = 0.01$).

Tab. 5 Instrument analysis (all 6-level items 1 = completely disagree, 6 = completely agree; N = 70) of the four affective outcome variables (interest (IN), self-concept (SC), curiosity state (CS) and relevance for science (RS) for pre and post data, and of the predictor variable curiosity as a trait (CT): number of items K, mean, item-test correlation $r_{it}$., Cronbach's $\alpha$; for the mean and $r_{it}$ the standard deviation SD and the [range of values] is given. Additionally, Cohen's $d$ for the pre/post comparison is reported (* indicates a small positive effect).

| Variable and example of item (translated from French) | Scale characteristics (SD, range) | PRE | POST |
|---|---|---|---|
| Interest (K = 5)<br>IN3: I liked the physics class. | Mean | 3.94(0.82)<br>[2.70 ; 4.56] | 4.08(0.92)<br>[2.97 ; 4.80] |
| | $r_{it}$ | 0.66(0.05)<br>[0.59 ; 0.72] | 0.73(0.01)<br>[0.72 ; 0.75] |
| | $\alpha$ | 0.65 | 0.77 |
| | Cohen's $d$ | +0.17 | |
| Curiosity state (K = 5)<br>CS4: The course aroused my curiosity about the topics covered. | Mean | 3.93(1.02)<br>[3.74 ; 4.20] | 4.40(0.98)<br>[4.00 ; 4.87] |
| | $r_{it}$ | 0.80(.05)<br>[0.75 ; 0.88] | 0.84(.08)<br>[0.72 ; 0.91] |
| | $\alpha$ | 0.87 | 0.89 |
| | Cohen's $d$ | +0.46* | |
| Self-concept (K = 5)<br>SC2: My classmates thought I am good at physics. | Mean | 3.76(0.99)<br>[3.27 ; 4.30] | 4.02(1.04)<br>[3.53 ; 4.43] |
| | $r_{it}$ | 0.80(0.04)<br>[0.74 ; 0.84] | 0.80(0.06)<br>[0.73 ; 0.88] |
| | $\alpha$ | 0.85 | 0.86 |
| | Cohen's $d$ | +0.26* | |
| Perceived relevance of science (K = 8)<br>RS2: Everyone should study the topics covered in the course. | Mean | 3.67(0.88)<br>[3.04 ; 4.47] | 3.68(0.89)<br>[2.60 ; 3.84] |
| | $r_{it}$ | 0.66(0.09)<br>[0.47 ; 0.75] | 0.65(0.07)<br>[0.52 ; 0.74] |
| | $\alpha$ | 0.81 | 0.80 |
| | Cohen's $d$ | +0.01 | |
| Curiosity trait (K = 5, as predictor)<br>CT1: In find fascinating to learn new things. | Mean | 5.03(0.48)<br>[4.53 ; 5.61] | - |
| | $r_{it}$ | 0.78(0.07)<br>[0.69 ; 0.87] | - |
| | $\alpha$ | 0.82 | - |

### 5.3 ANCOVA analysis

As for learning, ANCOVA analyses were conducted to examine the effects of gender, class (Tab. 1), prior physics grade, and curiosity trait (CT) on the variation of learning and affective outcomes (i. e. values post – values pre, see 4.3.3). No significant effects were found for any of the outcomes. Tab. 6 reports the *F* and *p*-values for all variables.

Tab. 6: The *F* and *p*-values of the ANCOVA analysis for the impact of predictros on the pre-post variation of the outcomes.

| Outcome post – outcome pre | Gender | Class | Prior physics grade | Curiosity trait |
|---|---|---|---|---|
| Curiosity state | $F = 0.09$<br>$p = 0.76$ | $F = 2.03$<br>$p = 0.12$ | $F = 0.12$<br>$p = 0.73$ | $F = 0.84$<br>$p = 0.36$ |
| Self-concept | $F = 1.95$<br>$p = 0.17$ | $F = 2.13$<br>$p = 0.11$ | $F = 2.28$<br>p = 0.14 | $F = 0.04$<br>$p = 0.84$ |
| Interest | $F = 0.87$<br>$p = 0.35$ | $F = 2.75$<br>$p = 0.05$ | $F = 0.08$<br>$p = 0.77$ | $F = 2.71$<br>$p = 0.10$ |
| Relevance science | $F = 2.17$<br>$p = 0.15$ | $F = 0.27$<br>$p = 0.85$ | $F = 1.08$<br>$p = 0.30$ | $F = 1.04$<br>$p = 0.31$ |
| Conceptual learning | $F = 0.88$<br>$p = 0.35$ | $F = 0.49$<br>$p = 0.69$ | $F = 0.09$<br>$p = 0.77$ | $F = 0.36$<br>$p = 0.55$ |

## 6 Discussion

### 6.1 Effects on learning (RQ1 and RQ2)

#### 6.1.1 Conceptual understanding

The results from section 5.1, namely the item and the instrument analysis of Tab. 2 and Tab. 3, supplemented by the interviews, allow us to positively address RQ1 and RQ2. In Tab. 3 we observe a significant increase in the rate of correct answers for each item, with effect sizes ranging from medium to large (Cohen's *d* between 0.4 and 1.6), as well as an overall Cohen's d of 2.8, indicating a remarkable overall effect. Observing the results of individual items, we can establish the following findings:

- The course's effect is likely strengthened by the low level of prior knowledge on these topics, which is evident from the small difficulty index ($p < 0.5$) for all items except for a couple, whose *p* value in the post-test is around 0.6: item 4 (on lensing) and item 12 (on black holes). Indeed, the items for which the effect size is “only” medium, show a relatively high *P* value already in the pre-test: this concerns item 4, item 8 (on the expansion's observational proofs), item 10 (on lensing), and item 12: all have a $P_{pre} > 0.4$. Nonetheless, the rate of correct answers after the course is eventually close to the average of the test ($P_{post\ average} = 0.87$, for item 10) or higher (items 4, 8, and 12).
- Item 5 (on dark matter) and 7 (on the nature of CMB) have a post $p > 0.9$, which reduces their discrimination coefficient *D* (to 0.18 and 0.38, respectively). However, the high *p* value indicates the success of the instruction for those concepts. In particular, the *D* value of item 5 is good in the pre-test, where the relative *p* value is within the allowed ranges.

- We observe the very low *P* value of item 11 in the pre-test. This reflects the conceptual difficulty in understanding that the density of a black hole is not necessarily high, and, on the contrary, for a supermassive black hole, it can be comparable to that of water or ordinary gas (see Supplemental Material [58]). However, after instruction, the *P*-value increases to approximately 40%.

### 6.1.2 Misconceptions

Beyond the simple analysis of binary results (right or wrong), in the pre-test, for 10 items out of 15, the wrong responses were not evenly distributed, and indeed the greater proportion of responses corresponded to one incorrect answer, revealing potential misconceptions. For each of those 10 items, Tab. 4 shows that all the potential misconceptions present in the pre-test are overtaken by the correct answers in the post-test. However, for certain questions, the answer corresponding to the misconception remains the second most common. In some cases, the misconceptions have been completely removed. In particular, after the instruction:

- only 6% (compared to 24% in the pre-test) of the students think that the observed speed of distant galaxies is about zero (item 2), static universe;
- no student claims that dark matter is composed of antimatter particles (item 5);
- only 6% of the students (compared to 37% in the pre-test) think that the CMB was emitted during the formation of the first galaxies (item 7);
- only 3% of the students (compared to 40% in the pre-test) think that dark energy is the energy associated with dark matter;
- less than 10% (compared to one-third in the pre-test) think that the universe is 150 billion years old.

Note that, for item 11, when grouping the pre-test responses that the density of a black hole 100 times more massive remains the same (answer C) or increases (whether by 100 or 10000 times, corresponding to D and E), we obtain 85% of responses. This percentage drops to 34% in the post-test, showing a strong reduction in the misconception that "a black hole is always very dense regardless of its mass". Similarly, grouping together the responses indicating that the density of a black hole 100 times more massive is less dense (whether it is 100 or 10000 times, i.e., responses A and B), they pass from 15% in the pre-test to 67% in the post-test. This means that, although 24% of students answered incorrectly in the post-test to this question by choosing response A (100 times less dense and not the correct answer of 10000 times), the misconception is eliminated for this group of students as well.

## 6.2 Effects on affective outcomes (RQ3)

Tab. 5 provides an answer to RQ3: it shows that all four affective dependent variables increased between the pre-test and post-test, and for two of them, this increase was significant, with a small to medium effect sizes: curiosity state (Cohen's $d = 0.5$) and self-concept in relation to the course (Cohen's $d = 0.3$). These quantitative findings are in line with the reasons discussed in sect. 2.2 likely to lead to a positive effect on the selected affective variables. We complement the discussion with some qualitative evidence that provides further insight from the semi-guided interviews (4.3.4).

First, student responses confirmed astronomy and astrophysics as "candidate topics" of high interest for young people. For example, one student remarked, "*I watch a lot of videos, I find cosmology and modern physics very fascinating*". Several other students also mentioned that they had already watched videos, read about topics like black holes and cosmic expansion, or explored these subjects in other ways, while recognizing that the level remained superficial compared to what was covered in the course. One student expressed satisfaction in having discussed in class the topic of supermassive black holes with an example that was featured in the news (the M87 black hole by the EHT, see Supplemental Material [58]): "*The* *course came at a very good time because the day after we covered the subject, we saw photos of the black hole in the center of our galaxy in the newspapers and were able to study it*". Moreover, the interviewed students unanimously appreciated the well-organized structure and the quality of the documents and materials (book and exercise sets), see e.g. the following statements: "*The book is incredible, excellent, it has everything in it*"; "*the documents were very precise and very well explained, the book is very well done, the files with summaries helped a lot*"; "*It was the most interesting course of the year, I loved it, I suggest that it always be integrated into the physics course*". This indicates that beyond the positive effect measured on learning, the level and structure are well adapted to the upper secondary school, answering to both RQ1 and RQ2.

Student feedback furthermore indicated a nuanced interaction between curiosity (CS) and interest (IN). In general, it is understandable that an in-depth treatment (relative to the students' level) of a certain content may not automatically increase interest in this content (overall interest level being relatively high, around 4 on a scale of 1 to 6). This can be explained as curious students (i.e., those "eager to learn new things") who are satisfied with their learning on a given subject prefer to learn new content related to what they have already studied, rather than dwelling on content they feel they have sufficiently covered for their level. For example, one student explained: "*I think that curiosity grew because the subject is truly interesting, and there's so much we don't yet know. But interest might decrease a bit because once you know the basics, the course inevitably gets more technical with math and physics. As for me, I like astronomy, but I wouldn't go further*". This potential dissociation of curiosity (for the topic) and interest (for learning in the course) is relevant for teaching practice, where one has to take additional measures in order to maintain sustained interest in learning.

Perceived relevance of science (RS) was relatively high in the sample (3.7 on a scale of 1 to 6), yet this variable showed no statistically significant increase, and the interviews confirm this result. We can nevertheless notice that, even if the course did not significantly change the students' view of scientific research, it can help students to better understand how research is conducted, the uncertain nature of scientific knowledge, and the usefulness of applying concepts learned in basic courses (accelerated or circular motion, inertia, free fall, the Doppler effect, …): "*I thought Newtonian mechanics was insufficient for calculating, for example, the Schwarzschild radius. I discovered that Newtonian mechanics can actually provide good approximations for more complex results*"; "*As for Newton's laws and waves, I can explain them now, and I feel more confident about these topics*".

A related observation is about students making connections and apply the concepts learned in the context of other disciplines, such as philosophy, realizing the interest of supporting certain philosophical concepts with data and quantitative observations within the framework of a scientific approach: "*It's actually the more philosophical side that is highlighted, thanks to the quantitative data, but also to my personal investigations*."

Overall, the students were very impressed by the scope of how much remains unexplainable in our universe, and some comments also testified to the impressions of "wonder" and "awe" mentioned in section 2.2 : "*I had never realized how many things we don't know: so many things that are still unknown to us!*"; "*Understanding that we are nothing compared to the universe, realizing that we are a tiny part of what surrounds us, was very striking to me*"; "*This allowed me to understand the notion of time and the limits of the physics of our world*"; "*I didn't think the temperature of the universe was that low. Our intuition is distorted by our senses and our daily experience*".

# 7 Conclusions

This study demonstrates that an effective cosmology course can be delivered at upper secondary level (RQ1) and highlights the significant potential for improving the knowledge of the students on these topics (RQ2). Furthermore, such a course leverages increased interest in these subjects to enhance their curiosity and possibly students' self-concept (RQ3), encouraging upper secondary students – both girls and boys in a balanced manner – to the world of modern physics research, which remains largely distant from the secondary school environment.

The following comment is in order: the project presented here aims at a coherent teaching-learning sequence of cosmology and general relativity, consistently maintaining a level accessible at high school. This does not mean, however, that the levels of abstraction and expertise required for these two topics are actually equal, or even similar; in fact, insights in cosmology (e.g. Hubble expansion) can be motivated using high school calculus and (pseudo-)Newtonian mechanics, but this is not the case for general relativity.

### 7.1 Perspectives

The complementary option course in this study was successful and, in the coming years, a sufficient number of students have enrolled for the annual version of the course (comprising two-semester modules instead of one). Starting in the fall of 2024, this optional course will also be offered in Italian at the Liceo of Lugano (Canton of Ticino), and the English translation will soon be available for a wider offer. In this context, the cosmology course materials used in this study are continuously updated, and improvements to the questionnaires (conceptual, affective, and interviews) are underway to continue evaluating their impact on the affective variables and understanding of upper secondary students. Specifically, the conceptual questionnaire will include questions on GR concepts as well. Concerning the affective variables, the participation of classes from different teachers and schools is planned, to distinguish the positive effect of the course from those arising from the context.

### 7.2 Creating new material

This course remains a challenge from the science education point of view, achievable only if appropriate educational materials, such as those described in section 3, are available and easily handled by teachers. The development and updating of the course materials for this study were made possible through extensive work (equivalent to 50% of a teacher's workload over several years plus 10% for the follow-up), involving continuous contact with both the research and teaching communities. Additionally, specific continuous professional education courses for teachers are necessary (see 7.3 below), in order to support for the newly developed teaching approaches and materials exchange with and implementation in the high school physics teaching community.

It is worth emphasizing that this effort cannot be considered part of the ordinary responsibilities of secondary school teachers, which typically involve selecting or adapting sequences or activities from existing materials and referring to predefined curricula and objectives. The specifications of a teacher do not include creating curricula (drawing up which subjects to cover in which year, at what level and, to some extent, in what order) or writing textbooks. This is evident in the traditionally taught subjects (mechanics, electromagnetism, thermodynamics, optics, etc.), for which predefined programs and materials exist.

For new content, where no sequence, reference level, program, or material exists, all of these elements must be created. This task is delicate, as the final work must serve as a foundation for teachers' responsibilities and requires a specific form of triple expertise: (i) a deep understanding of school environment (teachers and students); (ii) familiarity with research challenges in the new domain being taught (discipline research), and (iii) awareness of educational challenges at the target level (educational research).

Therefore, creating a complete course on a new subject area is a mission in itself and requires full institutional support.

### 7.3 Supporting the introduction of modern physics in schools

Furthermore, for this type of course to be truly offered and taught in classes, teachers must reasonably:

- Master the educational material;
- Feel comfortable conveying content to students that is inherently uncertain or incomplete.

Those two conditions constitute major obstacles to their engagement in teaching cosmology and, more generally, modern physics. Although we hope to have addressed here a milestone concerning the lack of adequate materials on cosmology, these two constraints remain non-trivial to meet. While the second point is inherent to the nature of science (NOS) and maybe the most valuable and fascinating aspect of teaching and studying these subjects at the frontier of knowledge, the first one is primarily a matter of insufficient resources available to support teachers. Indeed, instructors must possess a deeper understanding of the subject they teach than the level of knowledge they are expected to pass on to learners. To this end, the initial training process – aimed at enhancing the instructor's own knowledge, particularly regarding current research topics – represents a significant additional workload.
In this context, we recommend the following:

1. Modern physics topics should be integrated into the curricula, and the offer of complementary optional courses should be available at the upper secondary level.
2. Professional development courses and adjustments as workload reductions hours should be provided for teachers to support their readiness and encourage their engagement in teaching these subjects.

We hope this study has demonstrated the benefits and relevance of introducing cosmology at the upper secondary level.

## 8 Data availability statement

The raw data are available in the supplementary file data.csv [75] provided with this article.

## 9 Bibliography

[1] Klein, E. (2003). *Le temps existe-t-il?* Paris: Le Pommier.
[2] Bernardeau, F., Klein, E., & Laplace, S. (2013). *La physique des infinis*. Paris: La Ville Brûle.
[3] Barrau, A. (2018). *Trous noirs et espace-temps*. Bayard.
[4] Rovelli, C. (2018). *Reality is not what it seems*. Paperback.
[5] Turner, M. S. (2022). *The road to precision cosmology*. *Annual Review of Nuclear and Particle Science*, 72, 1–33.
[6] National Academies of Sciences, Engineering, and Medicine. (2021). *Pathways to discovery in astronomy and astrophysics for the 2020s*. Washington, DC: The National Academies Press.
[7] LIGO Caltech. (n.d.). Retrieved from https://www.ligo.caltech.edu
NASA Webb Space Telescope. (n.d.). Retrieved from https://webb.nasa.gov
NASA LISA. (n.d.). Retrieved from https://lisa.nasa.gov
SKAO. (n.d.). Retrieved from https://www.skao.int
[8] Michelini, M. (2021). Innovation of curriculum and frontiers of fundamental physics in secondary school: Research-based proposals. In B. G. Sidharth, J. C. Murillo, M. Michelini, & C. Perea (Eds.), *Fundamental physics and physics education research* (pp. 101–116). Springer International Publishing.
[9] UNESCO. (2011). *International standard classification of education: ISCED*. Paris: UNESCO.
[10] Brandt, S. (2009). *The harvest of a century: Discoveries of modern physics in 100 episodes*. Oxford University Press.
[11] Becchi, C. M., & D'Elia, M. (2016). *Introduction to the basic concepts of modern physics: Special relativity, quantum and statistical physics*. Cham: Springer.
[12] Butterfield, J., & Earman, J. (2007). *Handbook of the philosophy of science: Philosophy of physics A*. Amsterdam: Elsevier.

[13] Brush, S. G., & Belloni, L. (1983). *The history of modern physics: An international bibliography*. New York: Garland.
[14] Lewis, J. L. (1962). The teaching of modern physics in schools. *Physics Bulletin*, 13(1), 6.
[15] Ogborn, J. (1979). Modern physics curricula in higher education. *Reports on Progress in Physics*, 42(4), 727.
[16] Abell, S. K., & Lederman, N. G. (2007). *Handbook of research on science education*. Mahwah, NJ: Lawrence Erlbaum.
[17] Matthews, M. R. (2014). *International handbook of research in history, philosophy and science teaching*. Springer.
[18] Swiss Physical Society. (2022). Focus II: Impact of physics on Swiss society. *Communications of the Swiss Physical Society*. Retrieved from https://www.sps.ch/en/artikel/sps-focus/sps-focus-2
[19] Levrini, O. (1999). Teaching modern physics on the basis of a content knowledge reconstruction. In H. Behrendt et al. (Eds.), *Research in science education: Past, present, and future*. Proceedings of the Second International Conference of ESERA (pp. 301–312). Dordrecht: Kluwer.

[20] Gil, D., & Solbes, J. (1993). The introduction of modern physics: Overcoming a deformed vision of science. *International Journal of Science Education*, 15(3), 255–260.
[21] Pavlin, J., Stefanel, A., Lindenau, P., Kobel, M., Horvat, A. K., Wiener, J., & Čepič, M. (2021). Introduction of contemporary physics to pre-university education. In *JS21* (pp. 71–90).
[22] Zollman, D. (2016). Oersted lecture 2014: Physics education research and teaching modern physics. *American Journal of Physics*, 84(8), 573.
[23] Jarosievitz, B., & Sükösd, C. (Eds.). (2021). *Teaching-learning contemporary physics: From research to practice*. Springer International Publishing.
[24] Michelini, M., & Stefanel, A. (2023). Research studies on learning quantum physics. In *The international handbook of physics education research: Learning physics* (pp. 8-1). Springer.
[25] Gasparini, M. A. (2018). La cosmologie et la relativité générale par les mathématiques et la physique du lycée. *Communications of the Swiss Physical Society*, 55.
[26] Kersting, M., Henriksen, E. K., Bøe, M. V., & Angell, C. (2018). General relativity in upper secondary school: Design and evaluation of an online learning environment using the model of educational reconstruction. *Physical Review Physics Education Research*, 14(1), 010130.
[27] Gasparini, M. A., Müller, A., & Weiss, L. (2018). La cosmologie et la relativité générale au secondaire II au service de la motivation des élèves pour la physique. *Revue de Mathématiques pour l'école*, 230.
[28] Alstein, P., Krijtenburg-Lewerissa, K., & Van Joolingen, W. R. (2021). Teaching and learning special relativity theory in secondary and lower undergraduate education: A literature review. *Physical Review Physics Education Research*, 17(2), 023101.
[29] Weinberg, S. (1984). *The discovery of subatomic particles*. New York: Freeman.
[30] Kobel, M. (2003). High school students' exposure to modern particle physics. *Europhysics News*, 34(3), 108–110.
[31] Beck, H. P. (2021). Physics and education - Perspectives from particle physics. *Communications of the Swiss Physical Society*, 65, 22–24.
[32] Kranjc Horvat, A., Wiener, J., Schmeling, S. M., & Borowski, A. (2022). What does the curriculum say? Review of the particle physics content in 27 high-school physics curricula. *Physics*, 4(4), 1278–1298.
[33] Kragh, H. (2013). The science of the universe: Cosmology and science education. In *International handbook of research in history, philosophy and science teaching* (pp. 643–665). Dordrecht: Springer Netherlands.
[34] Salimpour, S., Fitzgerald, M., & Hollow, R. (2024). Examining the mismatch between the intended astronomy curriculum content, astronomical literacy, and the astronomical universe. *Physical Review Physics Education Research*, 20(1), 010135.
[35] Streit-Bianchi, M., Michelini, M., Binivento, W., & Tuveri, M. (2023). *New challenges and opportunities in physics education*. Cham: Springer.
[36] Kersting, M., & Woithe, J. (2022). IMPRESS: International Modern Physics & Research in Education Seminar Series. Retrieved from https://indico.cern.ch/category/15165/.

[37] Kersting, M., Blair, D., Sandrelli, S., Sherson, J., & Woithe, J. (2023). Making an IMPRESSion: Mapping out future directions in modern physics education. *Physics Education*, 59(1), 015501.
[38] Karplus, R. (1971). What is modern physics? *Physics Today*, 24(3), 13–14.
[39] Levrini, O., & Fantini, P. (2013). Encountering productive forms of complexity in learning modern physics. *Science & Education*, 22, 1895–1910.
[40] Baumert, J., Bos, W., & Watermann, R. (1998). *TIMSS/III – Schülerleistungen in Mathematik und den Naturwissenschaften am Ende der Sekundarstufe II im internationalen Vergleich*. Berlin: Max-Planck-Institut für Bildungsforschung.
[41] Levrini, O., & Fantini, P. (2013). Encountering productive forms of complexity in learning modern physics. *Science & Education*, 22, 1895–1910.
[42] Kragh, H. (2014). The science of the universe: Cosmology and science education. In *International handbook of research in history, philosophy and science teaching* (pp. 643–665). Dordrecht: Springer.
[43] Sjøberg, S., & Schreiner, C. (2007). Reaching the minds and hearts of young people: What do we know about their interests, attitudes, values, and priorities? What about the interest for space science? Bern: International Space Science Institute.
Sjøberg, S., & Schreiner, C. (2010). *The ROSE project: An overview and key findings*. Oslo, Norway: University of Oslo.
Lelliott, A., & Rollnick, M. (2010). Big ideas: A review of astronomy education research 1974–2008. *International Journal of Science Education*, 32(13), 1771–1799.
Baram-Tsabari, A., & Yarden, A. (2012). Characterizing children's spontaneous interests in science and technology. *International Journal of Science Education*, 27(7), 803–826.
[44] European Commission. (2021). *She Figures 2021 - Gender in research and innovation*. Luxembourg: Publications Office of the European Union. Retrieved from https://data.europa.eu/doi/10.2777/06090.
[45] Office Fédérale de Statistique (OfS). (2021). *OfS Numbers*. Retrieved from https://www.pxweb.bfs.admin.ch/pxweb/de/px-x-1502040100_103/.
[46] Hadzigeorgiou, Y. (2022). Students' reactions to natural and physical phenomena: Documenting wonder and engagement with science content knowledge. *Interdisciplinary Journal of Environmental and Science Education*, 18(1), e2261. https://doi.org/10.21601/ijese/11340.
[47] Valdesolo, P., Shtulman, A., & Baron, A. S. (2017). Science is awe-some: The emotional antecedents of science learning. *Emotion Review*, 9(3), 215–221.
[48] Yaden, D. B., Iwry, J., Slack, K. J., Eichstaedt, J. C., Zhao, Y., Vaillant, G. E., & Newberg, A. B. (2016). The overview effect: Awe and self-transcendent experience in space flight. *Psychology of Consciousness: Theory, Research, and Practice*, 3(1), 1–11.
[49] Galili, I., & Zinn, B. (2007). Physics and art: A cultural symbiosis in physics education. *Science & Education*, 16, 441–460.
[50] Barrau, A. (2023). *L'hypothèse K*. Bernard Grasset.
[51] Teaching Relativity. (n.d.). Retrieved from https://teaching-relativity.org/.
[52] Salinpour, S., Tytler, R., Doig, B., Fitzgerald, M. T., & Eriksson, U. (2023). Conceptualizing the cosmos: Development and validation on the cosmology concept inventory for high school. *International Journal of Science and Mathematics Education*, 21, 251–275.
[53] Possel, M. (2017). *The expanding universe: An introduction*. Lecture at the WE Heraeus Summer School "Astronomy from Four Perspectives," Heidelberg: Haus of Astronomie.
[54] Postiglione, A., & De Angelis, I. (2022). Introducing general relativity in high school: A guide for teachers. In *Challenges in Physics Education*. Springer, Cham.
[55] Gasparini, A. (2023). *Cosmologie et relativité générale, une première approche*. PPUR.
[56] Gasparini, A., & Müller, A. (2023). *Cosmologie et relativité générale: Activités pour les élèves du secondaire II*. Geneva.
[57] Gasparini, A., Stern, F., & Weiss, L. (2025). Un nouvel enseignement de la cosmologie et de la relativité générale expérimenté au gymnase. Submitted to *Progress in Science Education* (PriSE, CERN).
[58] See the Supplemental Materials file, containing more details on the course during the last part of the treatment duration, the arguments covered as well as the questionnaires administered: the conceptual learning test and the affective test.

[59] Ding, L., & Beichner, R. (2009). Approaches to data analysis of multiple-choice questions. *Physical Review Physics Education Research, 5*(2), 020103.
[60] Gasparini, A. (2021). *Using mobile devices as experimental tools in physics lessons: An empirical study of the effects on learning and motivation at secondary school level* (Ph.D. thesis). University of Geneva.
[61] Kuhn, J., & Müller, A. (2014). Context-based science education by newspaper story problems: A study on motivation and learning effects. *Perspectives in Science, 2*(1–4), 5–21.
[62] Hoffmann, L., Häußler, P., & Peters-Haft, S. (1997). *An den Interessen von Mädchen und Jungen orientierter Physikunterricht*. Christian-Albrechts-Universität, Kiel.
[63] Pawek, C. (2009). *Schülerlabore als interessefördernde außerschulische Lernumgebungen für Schülerinnen und Schüler aus der Mittel- und Oberstufe* (Ph.D. thesis). Christian-Albrechts-Universität Kiel.
[64] Marsh, H. W. (1990). The structure of academic self-concept: The Marsh/Shavelson model. *Journal of Educational Psychology, 82*(4), 623–636.
[65] Naylor, F. D. (1981). A state-trait curiosity inventory: Effect of computer graphics on improving estimates to algebra word problems. *Journal of Educational Psychology, 77*(3), 286–298.
[66] Litman, J. A., Collins, R. P., & Spielberger, C. D. (2005). The nature and measurement of sensory curiosity. *Personality and Individual Differences, 39*, 1123–1133.
[67] Van der Maren, J.-M. (1996). *Méthodes de recherche pour l'éducation*. Montréal, Québec: Presses de l'Université de Montréal.
[68] Cohen, J. (1988). *Statistical power analysis for the behavioral sciences* (2nd ed.). Lawrence Erlbaum Associates.
[69] Hattie, J. (2009). *Visible learning: A synthesis of over 800 meta-analyses relating to achievement*. Routledge.
[70] Benjamini, Y., & Hochberg, Y. (1995). Controlling the false discovery rate: A practical and powerful approach to multiple testing. *Journal of the Royal Statistical Society: Series B (Methodological), 57*(1), 289–300.
[71] Cronbach, L. J. (1951). Coefficient alpha and the internal structure of tests. *Psychometrika, 16*(3), 297–334. https://doi.org/10.1007/BF02310555
[72] R Core Team. (2021). *R: A language and environment for statistical computing*. Vienna: R Foundation for Statistical Computing. Retrieved from https://www.R-project.org.
[73] Kline, T. J. B. (2005). Classical test theory: Assumptions, equations, limitations, and item analyses. In *Psychological testing: A practical approach to design and evaluation* (pp. 107–136). SAGE Publications.
[74] Adams, W. K., & Wieman, C. E. (2011). Development and validation of instruments to measure learning of expert-like thinking. *International Journal of Science Education, 33*(9), 1289–1312.
[75] data.csv (2025). *Supplementary data for: Cosmology and general relativity in upper secondary school through new targeted teaching materials: a study on student learning and motivation*. Provided as supplementary material with the article in *Physical Review Physics Education Research*.

## Appendix 1. Course content covered during the last part of the intervention

After covering the introductory chapters on expansion and the basics of general relativity in the first half of the course, as described in Section 3, we briefly outline here the remainder of the course on gravitational lenses and black holes, referring to reference [Prise24] for more details.

### 1.1 Gravitational lensing

The next two chapters, namely the fourth on gravitational lensing and the fifth on black holes, focus on two remarkable applications of general relativity concepts. The teaching sequence on gravitational lensing begins with the derivation of the formula for the angle of deflection $\alpha$ (Fig. 1 on the left) of a light ray near a mass *M* through dimensional reasoning (without numerical factors) and then through the Newtonian approach – which differs from the relativistic formula by a numerical factor of ½. Beyond the numerical factor of this formula, the emphasis here is on the inverse dependence on the impact parameter: $\alpha \propto 1/d$, characterizing the gravitational lensing phenomenon. The comparison with the analogous situation of a converging optical lens, where the linear dependence on the distance between the incident ray and the optical axis ($\alpha \propto d$) implies the presence of an optical focus (Fig. 1 on the right), leads to questioning the shape of an optical lens that can reproduce this $1/d$ deflection, typical of gravitational lenses.

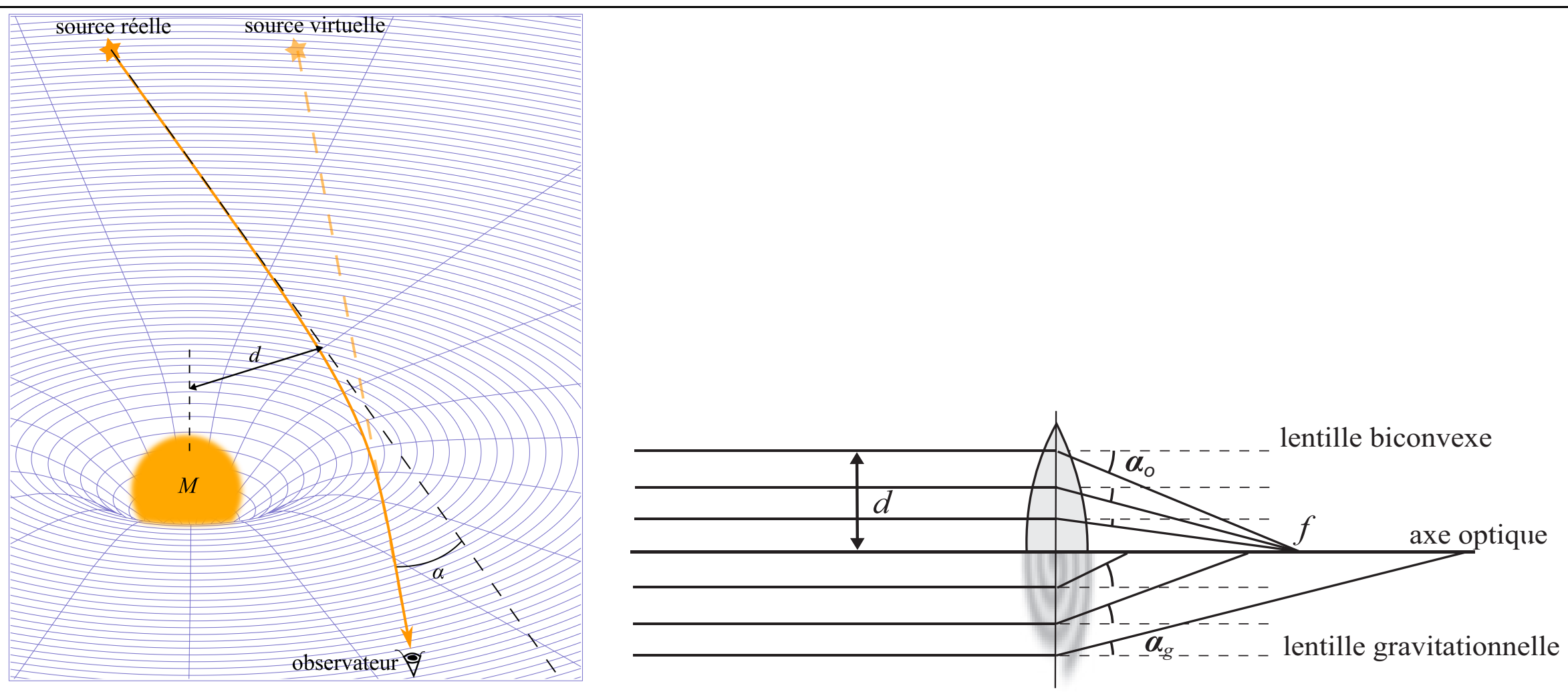

Fig. 1. Left: Diagram showing the angle of deflection of a ray passing near a mass *M*. Right: Comparison of the effects of gravitational and converging optical lenses on parallel rays.

The laws of refraction and a function integration are deployed to obtain this dependence: a logarithmic profile, similar to that of the bottom of a wine glass, is the one sought after (Fig. 2, left). However, since the students have previously learned that the curvature is negative around a mass with a spherically symmetric gravitational potential (**Error! Reference source not found.**), they can link the *1/d* dependence of the mathematical formula to the divergent behavior of parallel geodesics in the geometric representation. The manipulation visualizing the image of a point source through an optical lens of such shape is an activity that can easily be practiced in the classroom, creating the conditions for observing a ring, an Einstein cross, or gravitational arcs (Fig. 2, right). These images can then be compared to equivalent astrophysical images, for example, by Hubble or James Webb or the Event Horizon Telescope (EHT), but also with popular simulations such as the accretion disk ring of the black hole from the science fiction movie "Interstellar".

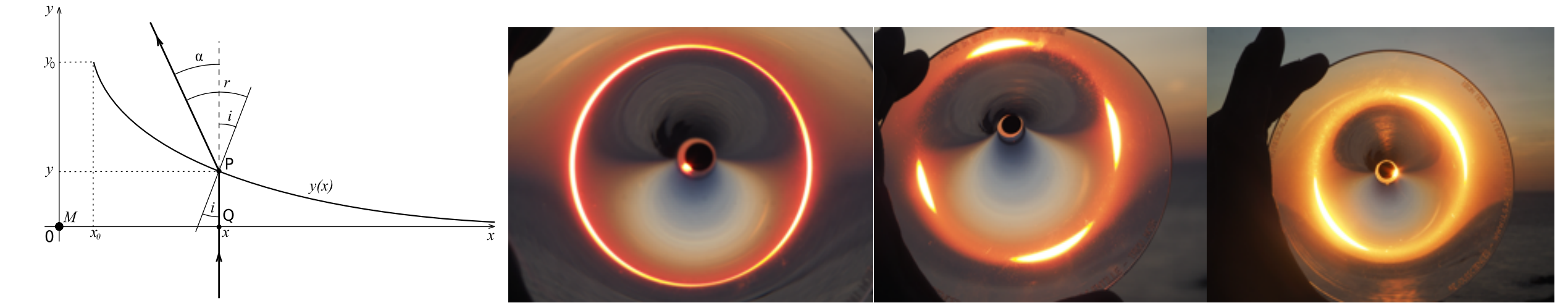

Fig. 2. Left: ray of light passing through an optical lens with a "bottom of a wine glass" profile. Right: image of a point-like source seen through a such a lens, showing a ring, an Einstein cross, or gravitational arcs.

In the case of perfect alignment between the observer O, lens L, and source S, the formula giving the Einstein radius $\theta_E$ as a function of the lens mass $M$ and the distances between the source and the lens is obtained from that of the deflection angle, using the law of sines and the "small angles" approximation. The formula thus obtained is used by astronomers to estimate the mass of the lens, including that of dark matter, as it is the gravitational mass of the lens that causes the phenomenon. The distances can be estimated from the redshifts of the source and lens, and the Einstein radius is a measurable quantity. Subsequently, exercises and activities related to these contents use real images and data from observations as a context for applying these conceptual and computational skills. While the most spectacular images are those related to strong gravitational lensing, the effects of microlensing and weak lensing are also addressed in the course, with a particular emphasis on the latter, which is fundamental in determining the proportion and distribution of dark matter – by nature less concentrated than baryonic matter. The main challenge here is to understand that research results primarily involve statistical analyses, putting into perspective the role of visual perception (but more generally, all human perceptions). The Euclid satellite, launched in the summer of 2023, has revealed numerous images with a significant media impact. However, the scientifically more interesting aspects cannot be grasped by simple human observation of these images. It is only by gaining a deeper understanding of the nature of the gravitational lensing phenomenon that one can appreciate the scope and challenges of this mission, sparking wonder and curiosity.

## 1.2 Black holes

Chapter 5 provides the ideal framework to introduce the concept of gravitational potential energy in $1/r$ (not present in the basics physics course in Switzerland). Then, applying the conservation of mechanical energy to the situation of an object launched from the surface of a celestial body allows deducing the formula for escape velocity, necessary for the Newtonian derivation of the Schwarzschild radius. Despite the derivation by Newtonian mechanics, students can refer to their knowledge acquired from the basics of the relativistic view of gravity and grasp the profoundly different interpretation of the concept of the event horizon that follows, as well as the reason why no signal can escape this boundary, as no geodesic exists allowing such a trajectory. The rigorous relativistic derivation is beyond the reach of high school students, but the geometric interpretation is conceptually within their grasp and allows for a deep understanding of the difference between the Einsteinian and Newtonian views of gravity. Thanks to its spherical symmetry, the situation is illustrated by a two-dimensional visualization of the spacetime structure inside the event horizon (Fig. 3).

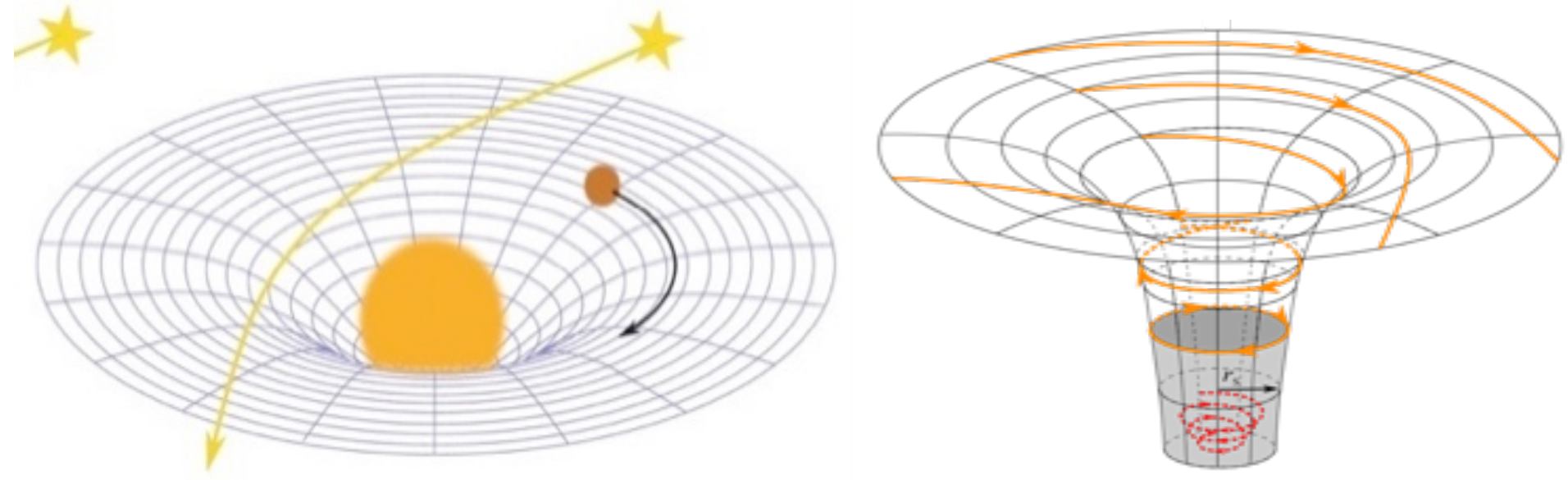

Fig. 3. Spacetime near a symmetric mass concentration: curvature is negative around the mass but becomes positive at the center and remains continuous for a star or a planet (on the left). In contrast, for a black hole, it remains negative and diverges at the center, where the discontinuity and singularity of the theory are located (on the right). Inside $r_S$, geodesics no longer emerge.

Beyond the conceptual aspects related to the Schwarzschild radius, it is also interesting to focus the mathematical relationship between the radius and the mass of a black hole: $r_S = \frac{2GM}{c^2}$. Firstly, the "small" (relative to our units) value of *G* in the numerator and the large value of $c^2$ in the denominator imply that the radius has a relatively small dimension compared to the masses of the most frequent astrophysical objects: for example, a black hole with the mass of Earth would have a radius of less than a centimeter. Furthermore, using Kepler's laws or the approximation of circular motion, students estimatie the mass inside the orbit of the star S02 around the radio source Sgr A* at the center of the Milky Way and compare its Schwarzschild radius with the perihelion of the star S17, leading to the conclusion that Sgr A* can only be a black hole, and going through the steps of the discovery awarded the Nobel Prize in 2020. The *direct* proportionality link between $M$ and $r_S$ has consequences that may seem surprising, especially due to the loss of intuition about the orders of magnitude involved when significantly departing from common-sense experience. Indeed, while one might initially think that having a relatively large mass in a relatively small region of space implies very high density, this reasoning does not apply to very massive black holes because, as $M$ and $r_S$ are proportional, if the volume of a black hole increases proportionally to the cube of its radius, this volume also increases with the cube of its mass. As a result, the average mass density of a black hole $\rho_{bh}$ is *inversely* proportional to the square of the mass – and thus the square of the radius – of the black hole:

$$V = \frac{4}{3}\pi R^3 \propto M^3 \quad \Rightarrow \quad \rho_{\rm bh} = \frac{M}{V} \propto \frac{M}{M^3} = \frac{1}{M^2}.$$

Indeed, after calculating some numerical mass densities, an algebraic replacement leads to the following formula $\rho_{\rm bh} = 3c^6/(32\pi G^3 M^2)$. While the numerical constants result in extremely high densities (relative to our units – for a mass of 1 kg, the mass density rises to approximately $\sim 10^{80}$ kg/m$^3$), the dependence on $1/M^2$ implies a decrease to the same order of magnitude as the densities observed in nature, around $10^{17}$ kg/m$^3$ for bodies of stellar mass (the smallest observed black holes), and "common" densities for masses of galactic scale, typically those of supermassive black holes in the nuclei of galaxies. Well-known examples include those of supermassive black holes pointed out by the EHT collaboration: M87*, whose image made headlines in 2019, with a mass on the order of a billion solar masses and a density of less than 1 kg/m$^3$, comparable to that of air. Students can calculate this density, and the image of the shadow and data related to M87* are used to distinguish between the Schwarzschild radius, the Einstein radius, and the "shadow" radius of the ring. The difference between these three radii is explained by the different paths of rays curved by the gravity of the black hole, as shown in Fig. 4. A mass density on the order of $10^6$ kg/m$^3$ can be calculated for the black hole of Sgr A*, whose image was published in 2020 by the same collaboration, which is an order of magnitude higher than that of the core of the Sun.

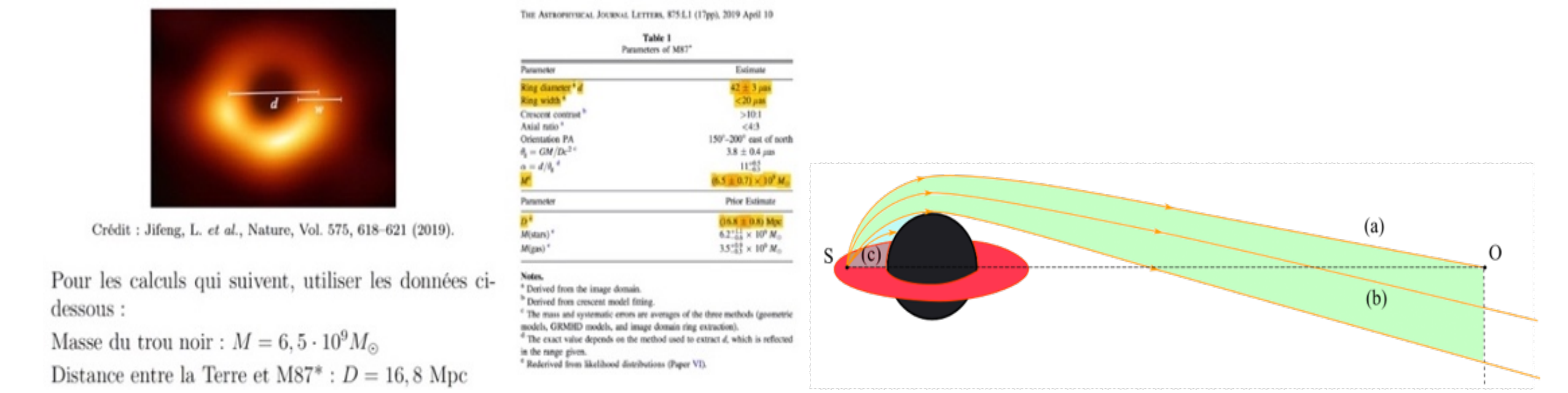

Fig. 4. The image of M87* and the related EHT data (on the left) are used in a course exercise. The diagram of the situation (on the right) with three types of rays originating from point S in the disk behind the singularity relative to the observer O.

Students then calculate the average density of a black hole with a mass equal to that contained within the Hubble radius ($M_H$), which coincides with the observed critical density of the universe: $\rho_{bh} \equiv \rho_c \sim 10^{-26} kg/m^3$. This result, demonstrated in a few steps, is a consequence of the definition of the "horizon" and is related to the finite value of the speed of light. Eventually, the dependence of the density of a black hole in $1/M^2$, shown in Fig. 5, is not unrelated to the tidal effect at its surface. After defining it as the derivative with respect to the radius of the acceleration of gravity, this effect is calculated near the surface of a black hole, also resulting in a shape in $1/M^2$. Consequently, the smallest black holes are not only the densest but also those with the greatest tidal effects for approaching bodies.

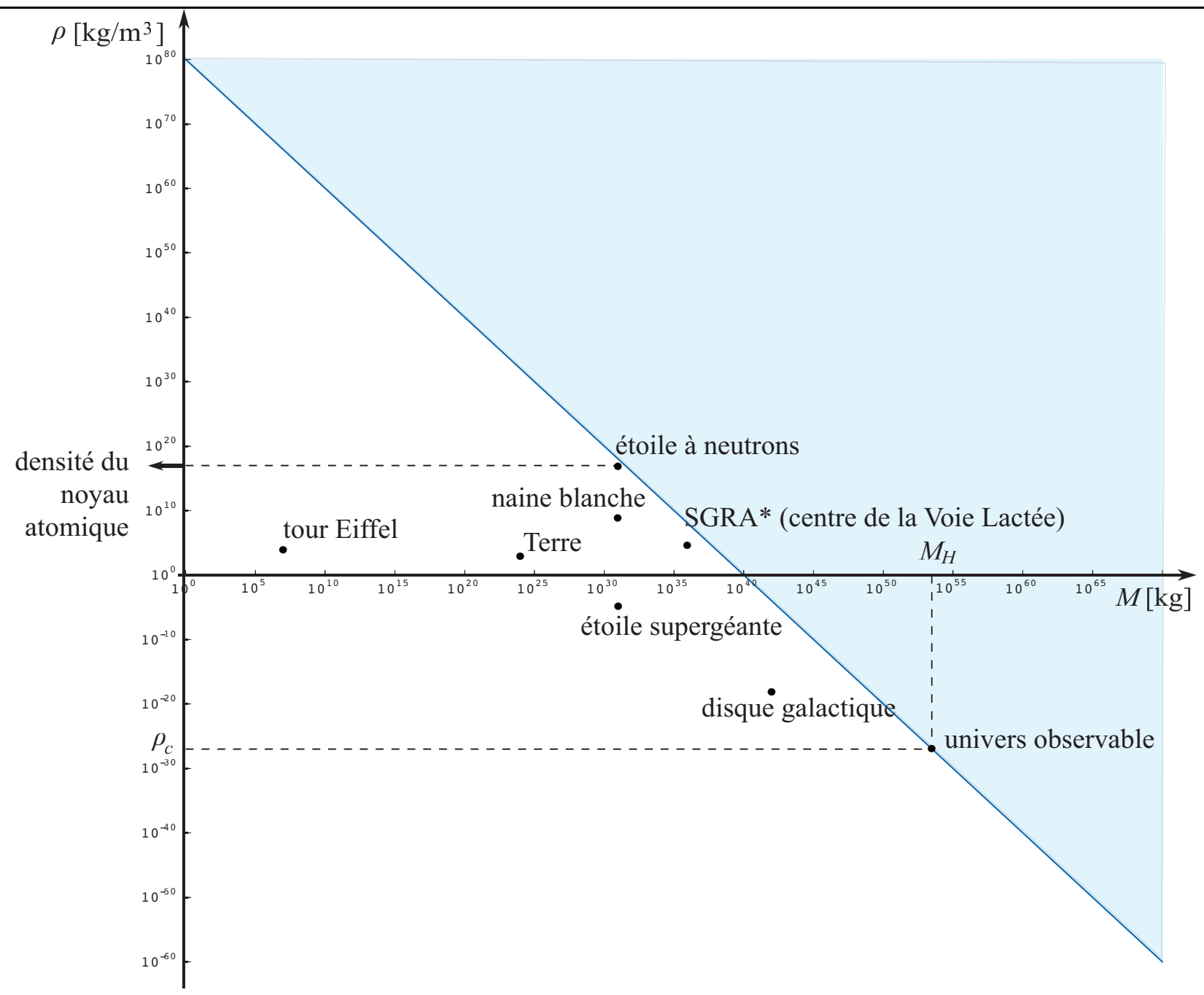

Fig. 6. The blue line represents, on an east logarithmic scale, the average density that a body would have to have to be a black hole, depending on its mass: $\rho_{bh} \propto M^{-2}$. The points representing few known objects are shown.

### 1.3 Temperature

The fact that a black hole is a radiating body and, therefore, has a nonzero temperature is a consequence of the inherently quantum nature of spacetime at microscopic scales. Understanding – even solely conceptually – the phenomenon of black hole evaporation provides an opportunity to apply the foundational concepts of both general relativity and quantum mechanics to a phenomenon that lies at the intersection of the validity domains of these two very different theories. Providing a rigorous and understandable explanation of this phenomenon at the high school level constitutes one of the most ambitious challenges of this course. The approach to the subject starts with the historical and macroscopic perspective of Bekenstein and Hawking, linked to the notions of entropy and the laws of thermodynamics. As in the case of the angle of deflection and the Schwarzschild radius, a simplified derivation yields a formula for the temperature as a function of the black hole's mass that differs only by a factor of 8π from Hawking's result, obtained with a rigorous derivation:

$$T = \frac{\hbar c^3}{8\pi k_B G M}$$

The derivation of the approximate formula is done through a semi-quantitative approach, starting from the "no-hair" theorem and Wien's law for the blackbody spectrum (also used to explain the connection between wavelength and temperature of the CMB). Although the mathematical level of this derivation is elementary, it involves a set of demanding conceptual and methodological knowledge from the scientific approach, such as those related to the concepts of heat transfer by radiation, blackbody spectrum, reasoning by exclusion, and order of magnitude. Firstly, as was the case for the derivation of the deflection angle of light, beyond the numerical factor of the formula, the emphasis is placed on the dependence of the equation on the parameters. The fact that the temperature is *inversely* proportional to the mass *M* (related to the radius), implies that the less massive a black hole is, the hotter it is, thus radiating and losing mass, becoming even hotter until complete evaporation. Thus, unlike what we observe for ordinary bodies – which, by radiating, lower their temperature until reaching an equilibrium state with their external environment – the phenomenon of black hole radiation is divergent. Secondly, similarly to what we have seen previously for the Schwarzschild radius and the average mass density (and for other quantities in the previous chapters), the numerical factors allow considerations about the orders of magnitude involved. Although $\hbar$ is "relatively" small ($\sim 10^{-34}$), the $c^3$ ($\sim 10^{25}$) in the numerator, as well as $k_B$ ($\sim 10^{-23}$) and *G* ($\sim 10^{-10}$) in the denominator, make the factor of this formula "relatively" large: ordinary mass black holes (on the order of kg or t) would have a significantly higher temperature than the average temperature of the universe, and, conversely, observed astrophysical black holes ($M > 10 M_{\odot}$) have temperatures much lower. Students can thus find what the limit mass of a black hole should be for it to radiate in the actual universe: around $10^{22}$kg, a hundred million times less massive than the Sun and a hundred times less massive than the Earth, corresponding to black holes never observed. These conclusions raise other questions specific to current research, for example: "Do such small black holes of such mass exist or have existed in the primordial universe?" or "Is every black hole destined to disappear with the cooling of the universe?" The answers to these questions remain largely unknown, but this course allows students to grasp the questions of research, aiming to increase their curiosity about these subjects, as we wille see in the section of the results.

After the macroscopic treatment through considerations related to thermodynamics, the semester concludes with an introduction to Heisenberg's uncertainty principle and the notion of virtual particle, foundational to the two microscopic interpretations of black hole evaporation:

- The interpretation originally given by Hawking represents quantum vacuum as a substrate of continuous creation and annihilation of virtual particle-antiparticle pairs. In this interpretation, the fact that smaller black holes radiate more intensely is explained by a more intense tidal effect at the horizon for small masses, as seen before.
- The interpretation through the tunneling effect: the quantum vision corrects the divergence of relativistic theory, considering that the wave function, and therefore the probability of detecting particles inside the black hole, cannot be concentrated at a single point due to indeterminacy, but rather has a bell-shaped distribution. The more intense emission from smaller black holes is explained here by a shorter distance separating the horizon from the center of the black hole. This interpretation also gives a new derivation of the black hole temperature formula, off by a factor of π compared to Hawking's. This is achieved by setting limits on the energy a particle can have outside $r_S$, due to the uncertainty principle, and deducing the associated temperature through energy equipartition.

# *Test de*

# *Astronomie et Cosmologie*

**Nom et prénom :**

**Date :**

**Groupe :**

**Enseignant·e :**

QCM : 14 questions

Durée : 20 minutes

- Chaque question comporte une seule réponse correcte.
- Chaque réponse correcte rapporte un point et il n'y a pas de déduction pour les réponses erronées.

# Bon travail !

1) Le Big Bang est

(A) une explosion à un instant précis dans le passé qui a donné naissance à l'univers.

(B) une explosion d'un astre massif qui s'est produite il y a plus de dix milliards d'années.

(C) l'instant dans le passé où le temps a commencé.

(D) un état de l'univers primordial que les théories physiques actuelles ne peuvent pas décrire.

(E) la phase du passé où l'univers était si chaud qu'il irradiait dans le visible.

2) Depuis la Voie lactée, nous pouvons observer les autres galaxies dans l'espace

(A) se rapprocher avec une vitesse d'autant plus intense que leur distance est grande.

(B) se rapprocher avec la même vitesse, quelque soit leur distance.

(C) s'éloigner avec une vitesse d'autant plus intense que leur distance est grande.

(D) s'éloigner avec la même vitesse, quelque soit leur distance.

(E) à peu près immobiles.

3) L'univers a-t-il un centre ?

(A) Non.

(B) Oui et il est à une très grande distance de la Terre.

(C) Oui et il se trouve dans la Voie lactée.

(D) Oui et il est en mouvement permanent.

(E) La question de l'existence ou non d'un centre de l'univers n'a pas encore été résolue.

4) Deux rayons de lumière parallèles passent à proximité d'un astre massif. Parmi les schémas suivants, lequel mieux décrit la trajectoire des deux rayons ?

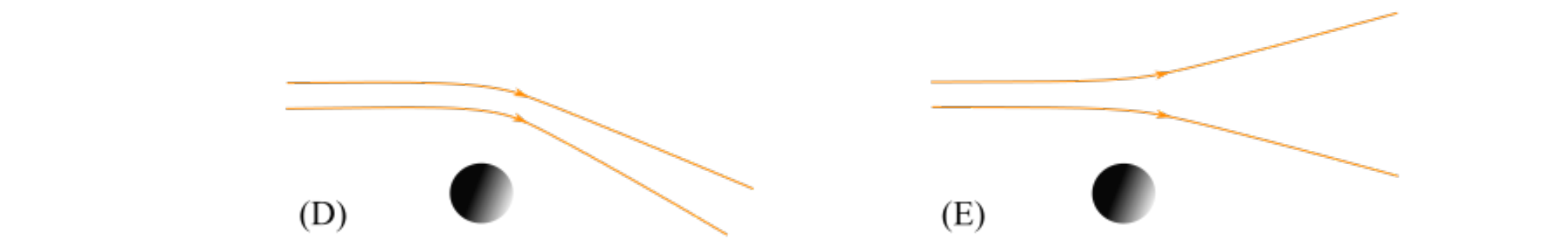

5) On appelle « matière noire »

(A) de la matière qui n'émet pas de radiation dans le visible.

(B) des particules d'antimatière.

(C) de la matière qui n'interagit pas électromagnétiquement.

(D) de la matière à l'intérieur des trous noirs.

(E) l'espace vide entre les galaxies.

6) Un signal lumineux part d'une galaxie en direction d'une autre galaxie. Au moment du départ du signal, les deux galaxies sont distantes l'une de l'autre de 700 millions d'années-lumière. Le signal lumineux

(A) met 700 millions d'années pour arriver d'une galaxie à l'autre.

(B) met moins de 700 millions d'années pour voyager d'une galaxie à l'autre, parce qu'entretemps la distance entre les galaxies a diminué en raison de la contraction de l'espace.

(C) met moins de 700 millions d'années pour voyager d'une galaxie à l'autre, parce que la gravité rapproche tous les objets les uns des autres au cours du temps.

(D) met plus de 700 millions d'années pour voyager d'une galaxie à l'autre, parce qu'entretemps la distance entre les galaxies a augmenté en raison de leur mouvement dans l'espace.

(E) met plus de 700 millions d'années pour voyager d'une galaxie à l'autre, parce qu'entretemps la distance entre les galaxies est augmentée en raison de l'expansion de l'espace.

7) Le « fond diffus cosmologique » est le rayonnement

(A) produit lors de la formation des premières galaxies dans l'univers primordial.

(B) émis dans l'univers primordial quand les atomes neutres se sont formés.

(C) émis lors de mécanismes qui produisent des substances radioactives dans l'univers.

(D) produit par des mécanismes de fusion nucléaire à l'intérieur du Soleil.

(E) émis par de la matière précipitant dans un trou noir.

8) Qu'est-ce qui est une preuve que l'univers a été très chaud et dense par le passé ?

(A) L'observation du mouvement des galaxies.

(B) La proportion de matière noire présente dans l'univers actuel.

(C) Le fait que la composition chimique de l'univers est actuellement d'environ 75% d'hydrogène et 25% d'hélium.

(D) La mesure de la densité de matière-énergie moyenne totale de l'univers.

(E) Il n'y a actuellement aucune preuve qui confirme cela.

9) Par « énergie noire » on entend

(A) l'énergie des ondes gravitationnelles.

(B) l'énergie à l'origine du changement de volume de l'espace vide.

(C) l'énergie de la matière noire.

(D) l'énergie des trous noirs.

(E) l'énergie qui a produit le Big Bang.

10) Considérez deux galaxies lointaines de forme sphérique, G1 et G2, parfaitement alignées par rapport à un observateur sur la Terre (O).

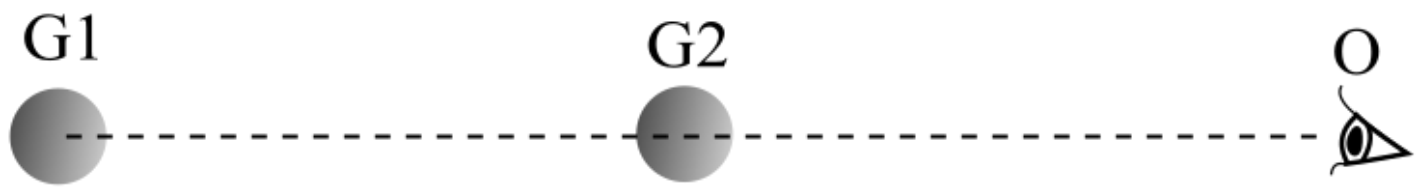

Dans cette configuration

(A) l'observateur (O) ne voit pas d'image de la galaxie la plus lointaine (G1).

(B) l'observateur (O) peut voir une image non déformée de la galaxie la plus lointaine (G1), déplacée par rapport à sa position réelle.

(C) l'observateur (O) peut voir deux images non déformées de la galaxie la plus lointaine (G1), déplacées par rapport à sa position réelle.

(D) l'observateur (O) peut voir une image déformée de la galaxie la plus lointaine (G1) dans sa position réelle.

(E) l'observateur (O) peut voir une image déformée de la galaxie la plus lointaine (G1), déplacée par rapport à la position réelle.

11) Considérez deux trous noirs, l'un avec une masse 100 fois plus grande que l'autre. Par rapport à la masse volumique du trou noir plus léger, la masse volumique du trou noir le plus massif est

(A) 100 fois plus petite.

(B) 10 000 fois plus petite.

(C) la même pour les deux trous noirs.

(D) 100 fois plus grande.

(E) 10 000 fois plus grande.

12) Un objet qui est tombé à l'intérieur d'un trou noir émet un signal sous forme de radiation électromagnétique en direction d'un observateur juste à l'extérieur du trou noir. L'observateur

(A) ne pourra jamais recevoir le signal émis.

(B) peut recevoir le signal très fortement décalé vers le rouge.

(C) peut recevoir le signal s'il attend assez longtemps.

(D) peut recevoir le signal seulement si la masse du trou noir est en dessous d'une certaine limite.

(E) peut recevoir le signal seulement si la masse du trou noir est en dessus d'une certaine limite.

13) Un trou noir isolé dans l'univers peut perdre de la masse ?

(A) Non, jamais. Un trou noir ne peut qu' « avaler » de la matière et donc augmenter sa masse.

(B) Oui, un corps constamment propulsé peut théoriquement s'échapper de l'intérieur d'un trou noir et donc contribuer à une perte de sa masse.

(C) Oui, si sa masse est en assez grande.

(D) Oui, si sa température est assez grande.

(E) Nous n'avons pas d'éléments pour savoir si un trou noir peut perdre de la masse.

14) Quelle est la meilleure estimation de l'âge de l'univers ?

(A) Environ 15 millions d'années.

(B) Environ 1,5 milliard d'années.

(C) Environ 15 milliards d'années.

(D) Environ 150 milliards d'années.

(E) L'univers a toujours existé et est donc infiniment âgé.

# ÉVALUATION DE LA MOTIVATION ACTUELLE DANS LES COURS DE PHYSIQUE

| Nom et prénom: | Date : |
|---|---|

## 1. Comment avez-vous trouvé le dernier module du cours OC de Physique ?

Avec ce questionnaire vous pouvez donner **votre avis**, concernant le cours de physique à option complémentaire **du module du <u>semestre passé</u>**. Veuillez mettre une croix sur la colonne qui correspond le mieux à votre opinion.

| | | **Avec cette affirmation, je suis…** | **pas d' accord du tout** | **pas d' accord** | **plutôt pas d' accord** | **plutôt d' accord** | **d' accord** | **tout à fait d' accord** |
|---|---|---|---|---|---|---|---|---|
| | 1 | J'aime apprendre des choses que je ne connais pas. | ① | ② | ③ | ④ | ⑤ | ⑥ |
| | 2 | J'ai bien aimé résoudre des problèmes proposés dans le dernier module du cours OCPY. | ① | ② | ③ | ④ | ⑤ | ⑥ |
| | 3 | Je pense que tout le monde devrait étudier les sujets traités au cours OCPY du dernier semestre. | ① | ② | ③ | ④ | ⑤ | ⑥ |
| | 4 | J'aime faire des recherches sur les choses que je ne comprends pas. | ① | ② | ③ | ④ | ⑤ | ⑥ |
| | 5 | J'ai eu des difficultés à comprendre les sujets traités dans le dernier module du cours OCPY. | ① | ② | ③ | ④ | ⑤ | ⑥ |
| | 6 | Le dernier module OCPY m'a rendu.e plus curieuse/eux à propos des phénomènes que nous ne pouvons pas encore expliquer. | ① | ② | ③ | ④ | ⑤ | ⑥ |
| | 7 | Le dernier module du cours OCPY m'a ouvert les yeux sur des métiers nouveaux et passionnants. | ① | ② | ③ | ④ | ⑤ | ⑥ |
| | 8 | Je trouve fascinant d'apprendre des nouvelles choses. | ① | ② | ③ | ④ | ⑤ | ⑥ |
| | 9 | Je pense que les sujets étudiés lors du dernier module OCPY vont améliorer mes chances de carrière. | ① | ② | ③ | ④ | ⑤ | ⑥ |
| | 10 | Je voudrais avoir le plus possible de physique à l'école. | ① | ② | ③ | ④ | ⑤ | ⑥ |
| | 11 | Je trouve fascinants les sujets traités dans le dernier module du cours OCPY. | ① | ② | ③ | ④ | ⑤ | ⑥ |
| | 12 | J'ai bien aimé le dernier module du cours OCPY. | ① | ② | ③ | ④ | ⑤ | ⑥ |
| | 13 | Le dernier module OCPY m'a rendu.e plus critique et sceptique. | ① | ② | ③ | ④ | ⑤ | ⑥ |
| | 14 | Je me suis investi.e davantage pour le dernier module du cours OCPYque pour les autres cours. | ① | ② | ③ | ④ | ⑤ | ⑥ |

| | | Avec cette affirmation, je suis… | pas d’ accord du tout | pas d’ accord | plutôt pas d’ accord | plutôt d’ accord | d’ accord | tout à fait d’ accord |
|---|---|---|---|---|---|---|---|---|
| | 15 | Je suis nul.le dans les sujets traités lors du dernier module du cours OCPY. | ① | ② | ③ | ④ | ⑤ | ⑥ |
| | 16 | Lorsque j’apprends quelque chose de nouveau, je veux en savoir davantage sur ce sujet. | ① | ② | ③ | ④ | ⑤ | ⑥ |
| | 17 | En plus des devoirs, j’ai consacré du temps libre aux sujets abordés dans le dernier module du cours OCPY. | ① | ② | ③ | ④ | ⑤ | ⑥ |
| | 18 | Je pense que, le dernier semestre, mes camarades ont trouvé que j’étais bon.ne au cours OCPY du dernier module. | ① | ② | ③ | ④ | ⑤ | ⑥ |
| | 19 | Je voudrais en savoir davantage sur les sujets traités dans le dernier module du cours OCPY. | ① | ② | ③ | ④ | ⑤ | ⑥ |
| | 20 | J’aime passer du temps à réfléchir sur les sujets traités dans le dernier module du cours OCPY. | ① | ② | ③ | ④ | ⑤ | ⑥ |
| | 21 | Je veux toujours examiner les choses en profondeur. | ① | ② | ③ | ④ | ⑤ | ⑥ |
| | 22 | J’ai pu résoudre les problèmes du dernier module du cours OCPY. | ① | ② | ③ | ④ | ⑤ | ⑥ |
| | 23 | J’aimerais approfondir les sujets traités dans le dernier module du cours OCPY. | ① | ② | ③ | ④ | ⑤ | ⑥ |
| | 24 | J’aimerais faire de la recherche en physique. | ① | ② | ③ | ④ | ⑤ | ⑥ |
| | 25 | J’ai bien compris les sujets traités dans le dernier module du cours OCPY. | ① | ② | ③ | ④ | ⑤ | ⑥ |
| | 26 | J’ai trouvé ennuyeux de résoudre les problèmes du dernier module du cours OCPY. | ① | ② | ③ | ④ | ⑤ | ⑥ |
| | 27 | Le dernier module OCPY a changé mon regard sur la nature. | ① | ② | ③ | ④ | ⑤ | ⑥ |
| | 28 | Le dernier module du cours OCPY a éveillé ma curiosité à propos des sujets traités. | ① | ② | ③ | ④ | ⑤ | ⑥ |